\documentclass[aps, pre, reprint, superscriptaddress,longbibliography, 10pt]{revtex4-2}
\usepackage{amsmath}

\usepackage{amsbsy}
\usepackage{amssymb}
\usepackage{graphicx}
\usepackage{color}
\usepackage{xcolor}

\usepackage{physics}
\usepackage{soul}
\usepackage{color}
\usepackage{bm}
\usepackage{array}
\usepackage{multirow}
\usepackage{verbatim}
\usepackage{natbib}
\usepackage{nccmath}
\usepackage{algorithm2e}
\RestyleAlgo{ruled}

\usepackage{amsthm}
\usepackage{mathrsfs} 
\usepackage{bbm}  
\usepackage{commath} 
\usepackage{comment}

\usepackage{hyperref}
\usepackage{cleveref}
\newcommand{\rhoT}{P}

\begin{document}
	
	\title{How target distributions shape optimal stochastic resetting}
	
	\author{Gregorio García-Valladares}
	\affiliation{Física Teórica, Multidisciplinary Unit for Energy Science, Universidad de Sevilla, Apartado de Correos 1065, E-41080 Sevilla, Spain}
	\author{Antonio Prados}
	\affiliation{Física Teórica, Multidisciplinary Unit for Energy Science, Universidad de Sevilla, Apartado de Correos 1065, E-41080 Sevilla, Spain}
	\author{Alessandro Manacorda}
	\thanks{Corresponding author: \href{mailto:alessandro.manacorda@roma1.infn.it}{alessandro.manacorda@roma1.infn.it}}
	\affiliation{CNR Institute of Complex Systems, UoS Sapienza, Piazzale A. Moro 5, 00185, Rome, Italy}
	\author{Carlos A. Plata}
	\thanks{C.A.P. and A.M. contributed equally to this work.}
	\affiliation{Física Teórica, Multidisciplinary Unit for Energy Science, Universidad de Sevilla, Apartado de Correos 1065, E-41080 Sevilla, Spain}
	
	\begin{abstract}
		
		We investigate the search of a target with a given spatial distribution in a finite one-dimensional domain. The searcher follows Brownian dynamics and is always reset to its initial position when reaching the boundaries of the domain (boundary resetting). In addition, the searcher may be reset to its initial position from any internal point of the domain (bulk resetting). Specifically, we look for the optimal strategy for bulk resetting, i.e., the spatially dependent bulk resetting rate that minimizes the average search time. The best search strategy exhibits a second-order transition from vanishing to nonvanishing bulk resetting when varying the target distribution. The obtained mathematical criteria are further analyzed for different monoparametric families of distributions, which sheds light on the properties that control the optimal strategy for bulk resetting. Our work paves new research lines in the study of search processes, emphasizing the relevance of the target distribution for optimal search strategies, and identifies a successful framework to address these questions.
		
	\end{abstract}
	
	\maketitle
	
	\section{Introduction}
	
	Nature is full of instances of first-passage processes~\cite{redner_book,metzler_book}, ranging from the first encounter of two reacting molecules to the first flower visited by a pollinator. Search processes can be formulated in terms of first-passage problems, with the focus on analyzing specific quantities related to the first-passage time---like the mean first-passage time (MFPT)---as key indicators of search performance. These problems are especially relevant from an applied perspective and have a significant impact across a wide range of disciplines, from biophysics---at various scales, from single molecules~\cite{Li_cells_11} to animal foraging~\cite{foraging_book,mendez_book}---to computing methods~\cite{optimal_search_book}, rescue mission strategies~\cite{Kratzke_rescue_10}, or reliability theory~\cite{reliability_book,DeBruyne_optimization_20}, to name just a few.
	
	The typical goal in this context is the optimization of the search strategy, e.g., finding out the strategy that minimizes the MFPT~\cite{Kusmierz_optimal_17}. Intermittent strategies that combine reactive random search phases with faster, nonreactive relocation phases, have proven to be highly effective to this end~\cite{Benichou_intermittent_11}. In its simplest implementation, stochastic resetting constitutes a specific instance of such intermittent strategies, with instantaneous relocations~\cite{Evans_resetting_11,Evans_applications_20}. 
	
	In the last years, optimization of stochastic resetting has been a prolific line of research~\cite{Evans_optimal_11,Evans_applications_20,Pal_time-resetting_16,Chechkin_renewal_18,Ahmad_transition_19,Besga_experimental-resetting_20,DeBruyne_optimization_20,Singh_confining_20,Faisant_experimental-resetting_21,DeBruyne_optimization_2021,DeBruyne_optimal_22,Garcia-Valladares_heterogenous_23,Sunil_profligacy_24}. The original simple proposal for stochastic resetting has been sophisticated by implementing more realistic reset mechanisms, e.g., using confining potentials~\cite{Pal_restart_17,Besga_experimental-resetting_20,Mercado-Vasquez_intermittent_20,Gupta_return_21,Faisant_experimental-resetting_21,Gupta_return_linear_21} or introducing some cost for the resets~\cite{Evans_refractory_2018,Maso-Puigdellosas_refractory_19,DeBruyne_optimization_20,Sunil_cost_2023,Olsen_cost_24,Garcia-Valladares_refractory_24}.
	
	Remarkably, most optimization problems in search processes have been posed for a target with a well-defined position, with very few exceptions~\cite{Kusmierz_optimal_17,Evans_applications_20,Garcia-Valladares_heterogenous_23,Evans_target_25}. In such setups, all the results, including optimal strategies, depend on the target position. Nevertheless, the usual situation in a realistic search problem is the searcher being unaware of the target position, for instance, a gatherer looking for food or a rescue mission searching for a lost victim. Therefore, the role of the degree of uncertainty about the target location---codified in its spatial distribution---is a significant issue, worth of being investigated to develop optimal strategies in realistic contexts. This has recently been explored by optimizing the distribution of times between resetting events~\cite{Evans_target_25}. 
	
	Moreover, it must be remarked that stochastic resetting has  predominantly been studied in infinite domains, where resets are always beneficial because they impede the trajectories drifting away from the target.  For finite domains, the literature is not so vast~\cite{Christou_bounded_15,Pal_interval_19,Durang_bounded_19,DeBruyne_optimization_20,Chen_bounded_22}. In different search setups, where details such as geometry or the number of targets differ, stochastic resetting has been observed to be detrimental in certain situations~\cite{Christou_bounded_15,Pal_interval_19,Durang_bounded_19}. In those cases, an optimal nonvanishing resetting rate was found only when the resetting location is close enough to the target position. Regimes where restart---at constant rate---is convenient to accelerate the search fulfill the celebrated CV criterion ~\cite{Reuveni_universal_16,Pal_restart_17}.
	
	When searches occur in a one-dimensional bounded domain, the advantage of resetting boundaries naturally arises: If the searcher reaches either wall of the box without finding the target, the optimal strategy is clearly to restart the search from its initial position. A typical illustration of stochastic resetting for search processes is how coming back to the starting point may improve the search of your keys at home. When introducing resetting boundaries, we argue that resetting is surely beneficial if the searcher has reached the attic of their house with no success. An analogous idea has been very recently introduced in search processes with multiple searchers~\cite{biswas25prl}.
	
	In this article, we address the gaps described above by analyzing how the target's spatial distribution in a finite domain impacts the optimal resetting strategy. To this end, we employ the paradigmatic model of a resetting Brownian searcher moving in a finite one-dimensional box. Therein, we will look for the best resetting strategy, allowing for it to be heterogeneous: The resetting rate depends in general on the position of the searcher. Our optimization is then different from the recently related work by Evans and Ray~\cite{Evans_target_25}, where the optimization is carried out over the distribution of times between reset events---keeping the resetting homogeneous. Space-dependent resetting has been previously considered in systems with a localized target~\cite{Evans_optimal_11}, and the stationary state has been obtained by using a quantum mechanical mapping~\cite{Roldan17pre}. On the one hand, we put our analytical focus on finding whether or not resetting expedites the search process depending on the target distribution. In this regard, we provide neat criteria for determining when bulk resetting is beneficial or detrimental. On the other hand, we illustrate how optimal resetting profiles can be determined for different specific setups.
	
	The rest of the article is organized as follows. In Sec.~\ref{sec-model}, the basics of the resetting Brownian searcher is introduced. We optimize the MFPT over homogeneous resetting rate in Sec.~\ref{sec:hom}. In Sec.~\ref{sec-het}, the resetting strategy is considered heterogeneous. Our focus is to determine whether the absence of bulk resetting constitutes a (local) minimum for the MFPT. Section~\ref{sec-num} is devoted to illustrate specific heterogeneous optimal resetting strategies. We put forward the main conclusions of our study as well as some interesting perspectives in Sec.~\ref{sec-concl}. Finally, some technical derivations and/or discussions can be found in the Appendices.
	
	\section{Model}
	\label{sec-model}
	
	We consider a Brownian searcher that freely diffuses in a one-dimensional box.
	
	Let $X(t)$ be its position at time $t$, and the searcher is looking for the target located at $X_T$. From a mathematical point of view, $X$ is a stochastic process and $X_T$ is a random variable with a certain distribution $\rhoT$. The distribution $\rhoT$ quantifies the degree of uncertainty about the location of the target; flat and single delta-peaked distributions, respectively, represent extreme cases of full uncertainty and full certainty. The target remains motionless in any trajectory, which has been referred to as a quenched-disorder target~\cite{Garcia-Valladares_heterogenous_23} due to the formal similarity with systems with quenched disorder~\footnote{As opposed to an annealed-disorder target, whose position would evolve in time.}. 
	
	The essential difference between a distributed target and distributed resetting positions---the latter is not considered in our analysis but has been previously studied~\cite{Evans_optimal_11,Evans_applications_20,Olsen_reset_distr_23,Mendez_reset_random_24,Toledo_reset_random_23}---is remarkable. In a single trajectory of the stochastic process, there is only one target drawn from its distribution in the former case, whereas there are multiple resetting locations, drawn from its corresponding distribution, in the latter case. We focus on the problem with a distributed target, motivated by its central role in search processes.
	
\begin{figure}          
		\includegraphics[width=0.48\textwidth]{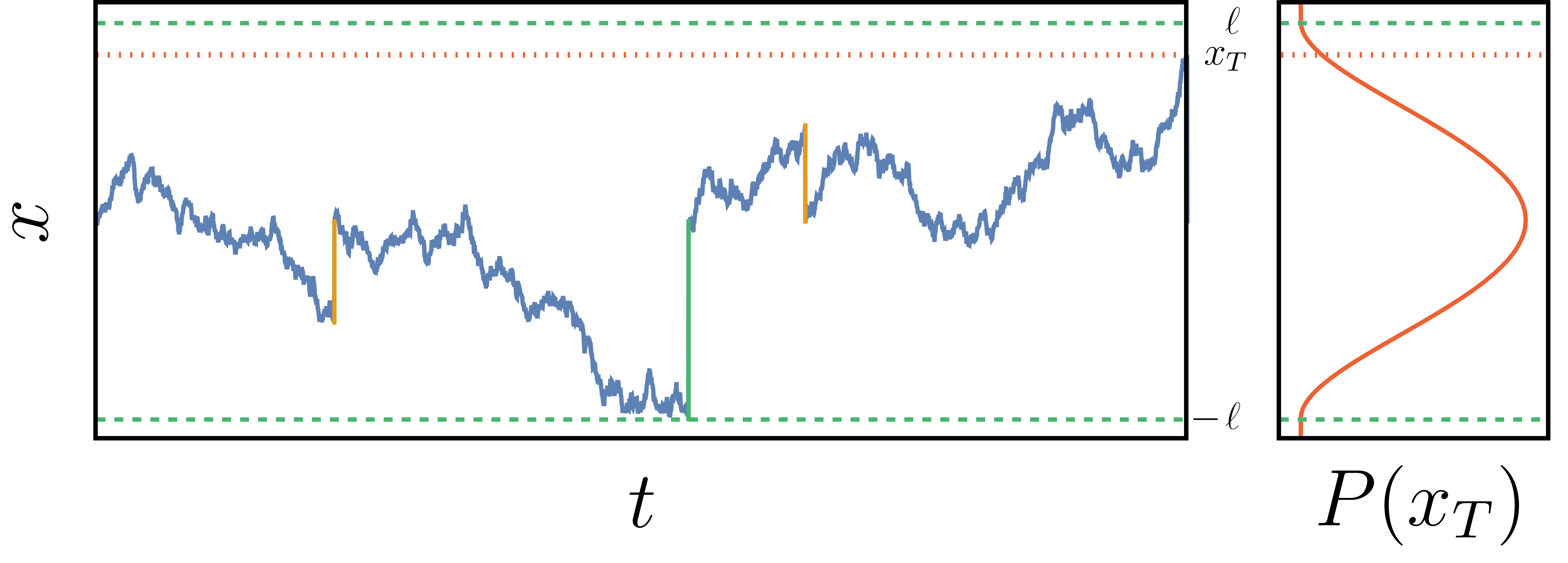}
		\caption{Sketch of a single trajectory for the Brownian search of a spatially distributed target. Left: One trajectory starting from $x=0$. The blue solid trace stands for diffusive motion, with instances of a boundary reset (vertical green stroke) and bulk resets (vertical yellow strokes). The target position $x_T$ (horizontal dotted red line) has been drawn from its distribution $\rhoT(x_T)$. Domain boundaries are at $\pm \ell$ (horizontal dashed green lines). Right: Spatial distribution of the target.}
		\label{fig:sketch}
	\end{figure}

	In our model dynamics, the searcher starts from the center of the box, $x=0$, for every trajectory of the stochastic process. When the searcher reaches the target at $x=x_T$, the trajectory terminates. Until then, dynamical rules---illustrated in Fig.~\ref{fig:sketch}---are as follows: 
	\begin{enumerate}
		\item When hitting the boundaries at $x=\pm \ell$, the position is instantaneously reset to the center;
		\item  the probability per unit time $r$ of a reset event in the bulk depends solely on space, i.e.,  $r\equiv r(x)$, and the reset event instantaneously restarts the search from the center;
		\item between reset events, the searcher freely diffuses with a constant diffusion coefficient $D$. 
	\end{enumerate}
	
	Hence, the equation governing the forward evolution of the probability distribution $p(x,t|x_0,t_0)$ of finding the searcher at position $x$ at time $t$, given that it was at position $x_0$ at time $t_0$ is
	\begin{align}
		\partial_t p(x,t|x_0,t_0) = &D \left\{ \partial_x^2 p(x,t|x_0,t_0) \right. \nonumber \\
		& \left. \qquad+ \delta (x) \left[ \partial_x p(x,t|x_0,t_0) \right]_{x=\ell}^{x=-\ell} \right\} \nonumber \\
		&- r(x) p(x,t|x_0,t_0) \nonumber \\
		&+ \delta(x) \int dx' r(x') p(x',t|x_0,t_0) ,
		\label{SMeq:forward1}
	\end{align}
	with initial condition
	\begin{equation}
		p(x,t_0|x_0,t_0)=\delta(x-x_0) ,\label{SM:forward2}
	\end{equation}
	and boundary conditions
	\begin{equation} \label{SM:forward3}
		p(x_T,t|x_0,t_0)=0 , \quad p(\pm \ell,t|x_0,t_0)=0.
	\end{equation}
	On the right hand side of Eq.~\eqref{SMeq:forward1}, the term in the first line stands for standard diffusion, the one in the second line stands for the gain of probability at $x=0$ stemming from the resetting at the boundaries. 
	The terms in the third and fourth lines respectively account for the probability loss at any $x$ and gain at $x=0$ after a reset in the bulk.
	
	Additionally, we have the initial condition for the propagator in Eq.~\eqref{SM:forward2}, the absorbing condition at the target position $x=x_T$ and the resetting boundary conditions at $x= \pm \ell$ in  Eq.~\eqref{SM:forward3}. Since the search instance is terminated when the searcher reaches $x=x_T$, for any search instance starting from $x_0=0$ the only active boundary is the one opposite to the target, namely $\pm \ell \to - \sigma(x_T) \ell$ in Eq.~\eqref{SM:forward3}, being $\sigma$ the sign function. The remaining condition would be always automatically satisfied~\footnote{Note that in a generic situation, both conditions (for $x=\pm \ell$) apply. More specifically, if we consider $x_T \, x_0>0$ and $0<|x_T|<|x_0|$, both conditions are required.}.
	
	On the one hand, our resetting boundaries clearly expedite---with respect to reflecting ones---the search in a one-dimensional domain: Reaching one wall without success means that the target lies on the other half of the box. This reasonable fact is checked for a specific setup at the beginning of Sec.~\ref{sec:hom} On the other hand, bulk resets at a certain position $x$ are beneficial (detrimental) if $x \, x_T$ is negative (positive), since resetting brings the particle closer to (farther from) the target. Therefore, two natural questions arise: 
	\begin{enumerate}
		\item When is bulk resetting advantageous for finding the target, given its distribution $\rhoT(x_T)$? 
		\item What is the optimal bulk resetting strategy $r(x)$ for a given $\rhoT(x_T)$? 
	\end{enumerate}
	The main goal of this study is to answer these two key questions.  
	
	Specific limits of the target distribution can already be discussed by relying on physical reasoning and relating them to previous results~\cite{Garcia-Valladares_heterogenous_23}. Specifically, if the target is certainly at a given position, the best heterogeneous resetting strategy is always to reset ($r(x) \to \infty$) for $|x_T-x|>|x_T|$ and to suppress resetting for $|x_T-x|<|x_T|$.   
	
	For a general search optimization, we start by studying the MFPT $\tau(x_0|x_T)$ from an arbitrary initial position $x_0$ to the target at a fixed location $x_T$. Specifically, we focus on the MFPT $\tau_0(x_T)\equiv \tau(0|x_T)$ from the center of the box to the target. Then, we consider its average with respect to the target distribution
	\begin{equation}
		\label{eq:av_tau}
		\overline{\tau} [r] \equiv \expval{\tau_0(x_T)}_T= \int dx_T \,\tau_0(x_T) \rhoT(x_T),
	\end{equation}
	where we have introduced the notation $\langle \cdot \rangle_T$ for averages with respect to $\rhoT(x_T)$. In this study, the average time $\overline{\tau} [r]$ is the main figure of merit for quantifying the success of the search. Note that $\overline{\tau} [r]$ is a functional of the bulk resetting strategy $r(x)$ through $\tau_0$---the functional dependence of the latter on $r(x)$ has not been explicitly included in the notation to avoid cluttering the formulas. 
	
	The equation governing the first-passage problem is the backward version of Eq.~\eqref{SMeq:forward1}. The MFPT at fixed $x_T$ is determined by
	\begin{subequations}
		\label{eq:taus}
		\begin{align}
			\partial_{x_0}^2 \tau(x_0|x_T)&=-1 +r(x_0) [\tau(x_0|x_T)-\tau(0|x_T)] ,\label{BFP_bulk}\\
			\tau(\pm 1 | x_T)&=\tau(0|x_T), \qquad \tau(x_T|x_T)=0, \label{BFP_boundary}
		\end{align}
	\end{subequations}
	where we have switched to dimensionless variables, respectively taking $\ell$ and $\ell^2/D$ as  units for space and time. The boundary conditions~\eqref{BFP_boundary} implement resetting boundaries at the walls and the absorbing point at the target position, respectively.
	
	Since we are interested in $\tau_0(x_T)$, it suffices to solve Eq.~\eqref{eq:taus} for $x_0$ in the interval between one of the walls and the target position that includes $x_0=0$, that is,  either $[-1,x_T]$ for $x_T>0$ or $[x_T,1]$ for $x_T<0$---analogously to the situation for the boundary conditions for the probability distribution, as discussed after Eq.~\eqref{SM:forward3}.
	
	Equation~\eqref{eq:taus} is simplified with the change of variables 
	\begin{equation}
		F(x_0)=\tau(x_0|x_T)-\tau(0|x_T),
	\end{equation} which removes the source term $\tau(0|x_T)$,
	\begin{subequations}
		\label{eqF-compl}
		\begin{align}
			-1 &= \partial_{x_0}^2 F(x_0) - r(x_0)  F(x_0) \ , \label{SM:eqF} \\
			F(0)&=0, \quad F(\pm 1) = 0 \ . \label{SM:eqF-bc}
		\end{align}
	\end{subequations}
	The dependence on $x_T$ is not explicitly written from now on to make our notation more concise. Since we are interested in the MFPT from the origin to the target position, once we solve the differential problem for $F$, we need to evaluate the solution at $x_0=x_T$, i.e., $\tau_0(x_T)=-F(x_T)$.  As for Eq.~\eqref{SM:forward3}, the boundary conditions~\eqref{SM:eqF-bc} make the system overconstrained; although they are both valid, Eq.~\eqref{eqF-compl} will be solved only by activating the boundary condition corresponding to $x_0=-\sigma(x_T)$.  While the analytical solution for any bulk resetting rate $r(x)$ is out of reach, it can still be obtained for specific choices of $r(x)$.
	
	\section{Optimization over homogeneous resetting}
	
	\label{sec:hom}
	We here consider the average MFPT defined in Eq.~\eqref{eq:av_tau} for a homogeneous bulk resetting rate $r$. Since resetting boundaries have been proved to be beneficial for the search problem, we will always assume their presence---even when bulk resetting vanishes, i.e., $r=0$. The average MFPT is then no longer a functional, but a standard function of $r$---we thus write $\overline{\tau}(r)$. We will optimize $\overline{\tau}(r)$ for a given $\rhoT$, i.e., we look for the best homogeneous resetting rate $r^*$ that minimizes $\overline{\tau}$. As already stated above, the dependence on $r$ enters through the solution for $\tau_0$ from Eq.~\eqref{eq:taus}, which can be solved for constant $r$, 
	\begin{equation}
		\tau_0(x_T;r)=r^{-1}\left[\frac{\cosh[(|x_T|+\frac{1}{2})\sqrt{r}]}{\cosh[\sqrt{r}/2]}-1\right] \ . \label{eq:sol-r-const}
	\end{equation}
	
	Had we consider reflecting boundaries instead of our resetting boundaries, the result would have been
	\begin{equation}
		\tau^{\text{(refl)}}_0(x_T;r)=r^{-1}\left[\frac{\cosh[(|x_T|+1)\sqrt{r}]}{\cosh[\sqrt{r}]}-1\right] \ , \label{eq:sol-r-const-refl}
	\end{equation}
	where we have taken into account that boundary conditions have changed. Specifically, the MFPT still verifies Eq.~\eqref{SM:eqF}, but the boundary conditions for $x_0=\pm1$ in Eq.~\eqref{SM:eqF-bc} are substituted with $\left. \partial_{x_0} F(x_0) \right|_{x_0=\pm1}=0$. As physically reasonable, it is straightforward to check that $\tau^{\text{(refl)}}_0(x_T;r) \geq \tau_0(x_T;r)$, $\forall (x_T;r)$: This constitutes a mathematical proof of the advantage of resetting boundaries over reflecting ones. Henceforth, we focus on the setup with resetting boundaries.
	
	To obtain $\overline{\tau}(r)$, it suffices to compute the average of Eq.~\eqref{eq:sol-r-const} with respect to $\rhoT(x_T)$. We note that $\tau_0(x_T;r)$ is a convex function of $r$, $\partial_r^2 \tau_0(x_T;r)>0 $, which diverges for $r\to\infty$. Convexity is inherited for the average MFPT $\overline{\tau}(r)$ for arbitrary $\rhoT(x_T)$. Since $r \geq 0$ by definition, the behavior of $\overline{\tau}(r)$ close to $r=0$ depends on the sign of $d\overline{\tau}/dr|_{r=0}$. For positive derivative, the global minimum of $\overline{\tau}(r)$ is located at $r=0$---due to the already discussed convexity: Therefore, the best search strategy implies no resetting in the bulk, $r^*=0$. Instead, for negative derivative, the global minimum of $\overline{\tau}(r)$ is no longer located at $r=0$: There appears an optimum nonzero resetting rate, $r^*\ne 0$; recall that $\overline{\tau}(r)$ diverges for $r \to \infty$. 
	
\begin{figure}
		
		\centering
		
		\includegraphics[width = 0.49\textwidth]{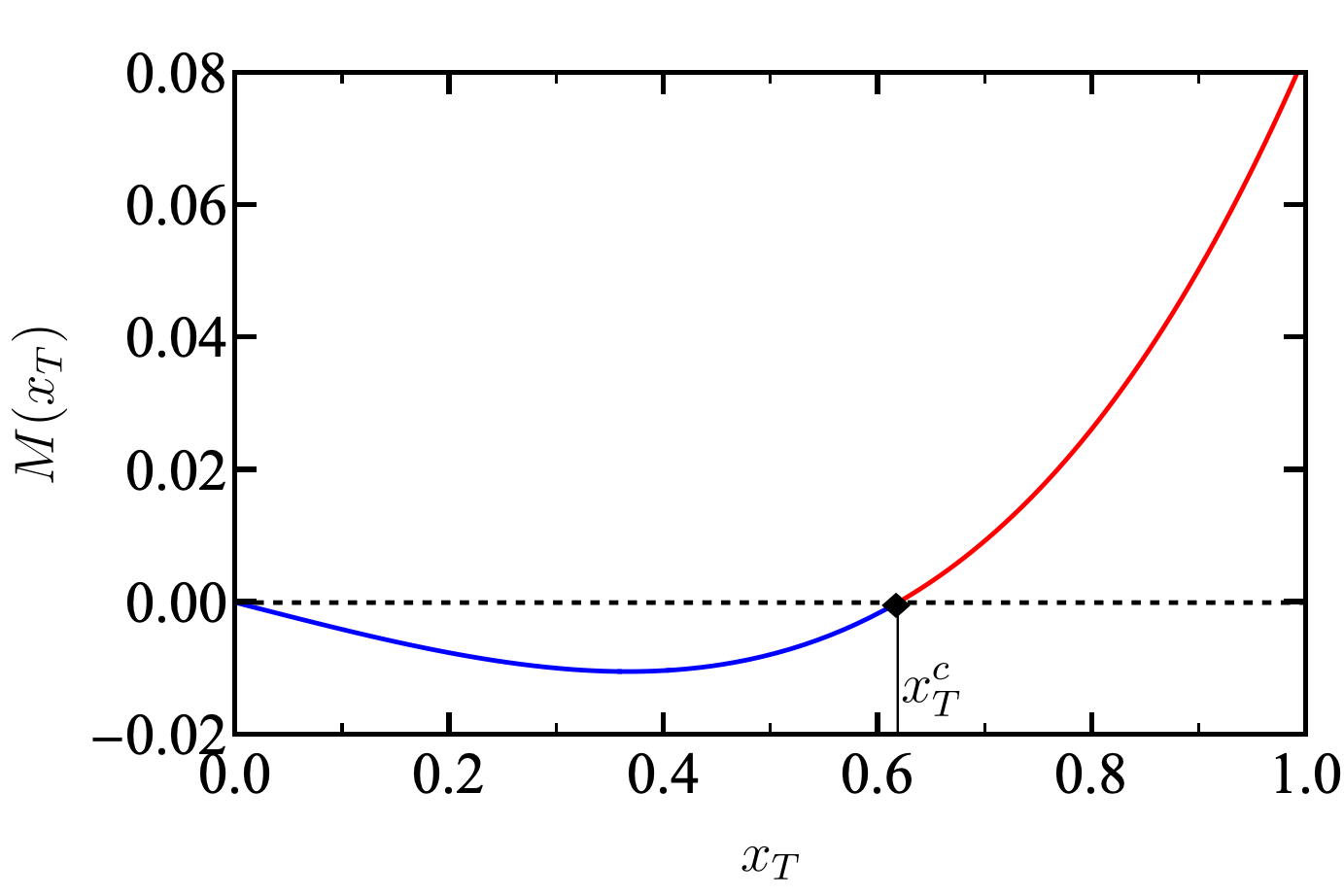}
		
		\caption{Derivative controlling the stability of the nonresetting strategy for homogeneous resetting, i.e., $M(x_T)$ defined in Eq.~\eqref{eq:m}. Positive (negative) values, represented as warm (cold) colors, thereof contribute to zero (nonzero) bulk resetting as the best search strategy. The change of stability is reached at $x_T^c$ (black diamond). }
		\label{fig:stability-hom}
	\end{figure}

	The above discussion implies that the key quantity is
	\begin{equation}\label{eq:m-def}
		m\equiv \left.\dv{\overline{\tau}}{r}\right|_{r=0}= \expval{M(x_T)}_T, \quad M(x_T)\equiv \left.\pdv{\tau_0(x_T;r)}{r}\right|_{r=0}.
	\end{equation}
	Specifically, $m>0$ ($m<0$) entails that the minimum of $\overline{\tau}$ is attained at $r^*=0$ ($r^*>0$). The analysis of the slope at $r=0$ has been successfully used in the past for optimizing different resetting configurations~\cite{Christou_bounded_15,Ahmad_transition_19,Ray_Peclet_19,Pal_Landau_19}.   
	Using Eq.~\eqref{eq:sol-r-const}, we get 
	\begin{equation}\label{eq:m}
		M(x_T)=\frac{1}{24}\left[x_T^4 - \abs{x_T} \left( 1-2x_T^2\right)\right].
	\end{equation}
	As shown in the top panel of Fig.~\ref{fig:stability-hom}, $M(x_T)<0$ for $|x_T|<x_T^c\equiv (\sqrt{5}-1)/2$, whereas $M(x_T)>0$ for $x_T^c<|x_T|<1$. Therefore, ``small'', close to the center, values of $x_T$ contribute to destabilize the nonresetting strategy in the bulk, whereas ``large'', close to the walls, values of $x_T$ contribute to stabilize it. 
	
	It is worth noting some general results. First, for a flat distribution, $m>0$; in a finite domain, homogeneous resetting does not give any advantage for a homogeneously distributed target. Second, $m<0$ for any target distribution with finite support in the subinterval $0\le x_T< x_T^c $,  because $M(x_T)<0$ where $\rhoT(x_T)\ne 0$: Therein, homogeneous resetting in the bulk is advantageous and $r^*\ne 0$. Conversely, $m>0$ for any target distribution with finite support in the subinterval $x_T^c<  x_T< 1 $, because $M(x_T)>0$ where $\rhoT(x_T)\ne 0$: Therein, homogeneous resetting in the bulk is detrimental and $r^*=0$. The discussion is analogous for $-1<x_T<0$ since Eq.~\eqref{eq:m} is an even function of $x_T$.
	
	To further illustrate our discussion, we now introduce a specific monoparametric family of symmetric target distributions. In particular, we choose a variant of the beta distribution,  
	\begin{equation}
		\label{eq:PTbeta}
		\rhoT(x_T;\beta)=\frac{\Gamma\left(\beta+\frac{1}{2}\right)}{\sqrt{\pi} \Gamma(\beta)}(1-x_T^2)^{\beta-1},
	\end{equation}
	defined for $-1 < x_T < 1$, with $\beta >0$.  This $\beta$-family interpolates between a single delta peak at the center of the box  for $\beta \to +\infty$ and two equally  weighted delta peaks at the boundaries for $\beta \to 0$, passing through the flat distribution for $\beta=1$. Since its support is the whole domain $[-1,1]$, the general observations made above for  especially supported distributions do not apply because there is a competition between positive and negative contributions to $M(x_T)$. Introducing Eqs.~\eqref{eq:sol-r-const} and~\eqref{eq:PTbeta} into Eq.~\eqref{eq:av_tau}, it is possible to analytically obtain the average MFPT, as detailed in Appendix \ref{Ap-hom}. 
	
	Yet, for our discussion, it suffices to obtain the linear coefficient $m$ in Eq.~\eqref{eq:m-def}, which is now a function of $\beta$:
	\begin{align}
		m(\beta)=\frac{1}{8[3+4\beta(\beta+2)]}-\frac{(\beta-1)\Gamma(\beta +\frac{1}{2})}{24\sqrt{\pi} \Gamma(\beta+2)}.
		\label{eq:m-hom-beta}
	\end{align}
	We note that $m(\beta)$ only vanishes at $\beta=\beta_c \simeq 1.71$. For $\beta<\beta_c$, we have $m>0$ and the global minimum of $\overline{\tau}(r)$ is found at $r=0$. Instead, $m<0$ for $\beta>\beta_c$.
	
	Therefore, the global minimum of $\overline{\tau}(r)$ is attained at a certain value $r^*>0$. This behavior is reminiscent of a continuous phase transition, with the optimal homogeneous resetting rate $r^*$ playing the role of the order parameter: $r^*=0$ for $\beta<\beta_c$,  $r^*\ne 0$ for $\beta>\beta_c$, with $r^*=0$ for $\beta=\beta_c$.
	
	For the homogeneous case, we have proven that nonresetting in the bulk is the optimal strategy for certain target distributions, specifically those for which the region with positive $M(x_T)$ (close to the walls) is more important, when weighted with the target distribution $\rhoT(x_T)$, than the region with negative $M(x_T)$ (close to the center). For the monoparametric family~\eqref{eq:PTbeta}, this corresponds to $\beta<\beta_c$: Loosely speaking, the weight of targets close to the center is insufficient to destabilize the nonresetting strategy in the bulk. Indeed, this is physically sensible: Since the resetting boundaries already facilitate the search, extra resetting in the bulk is beneficial only if the weight of targets close to the center is large enough.
	
	\section{Optimization over heterogeneous resetting}
	\label{sec-het}
	\begin{figure}
		
		\centering
		
		\includegraphics[width = 0.48\textwidth]{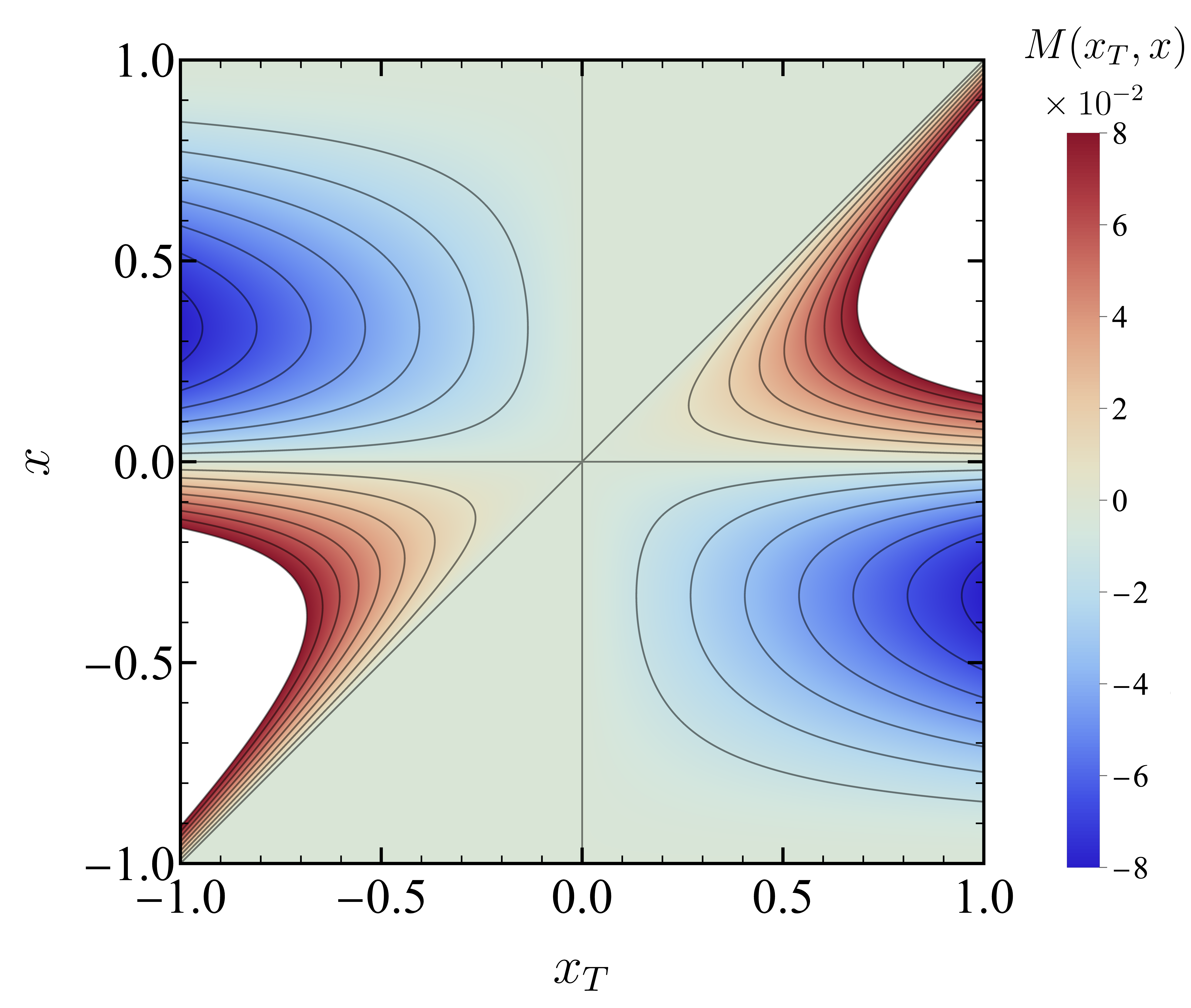}
		\includegraphics[width = 0.48\textwidth]{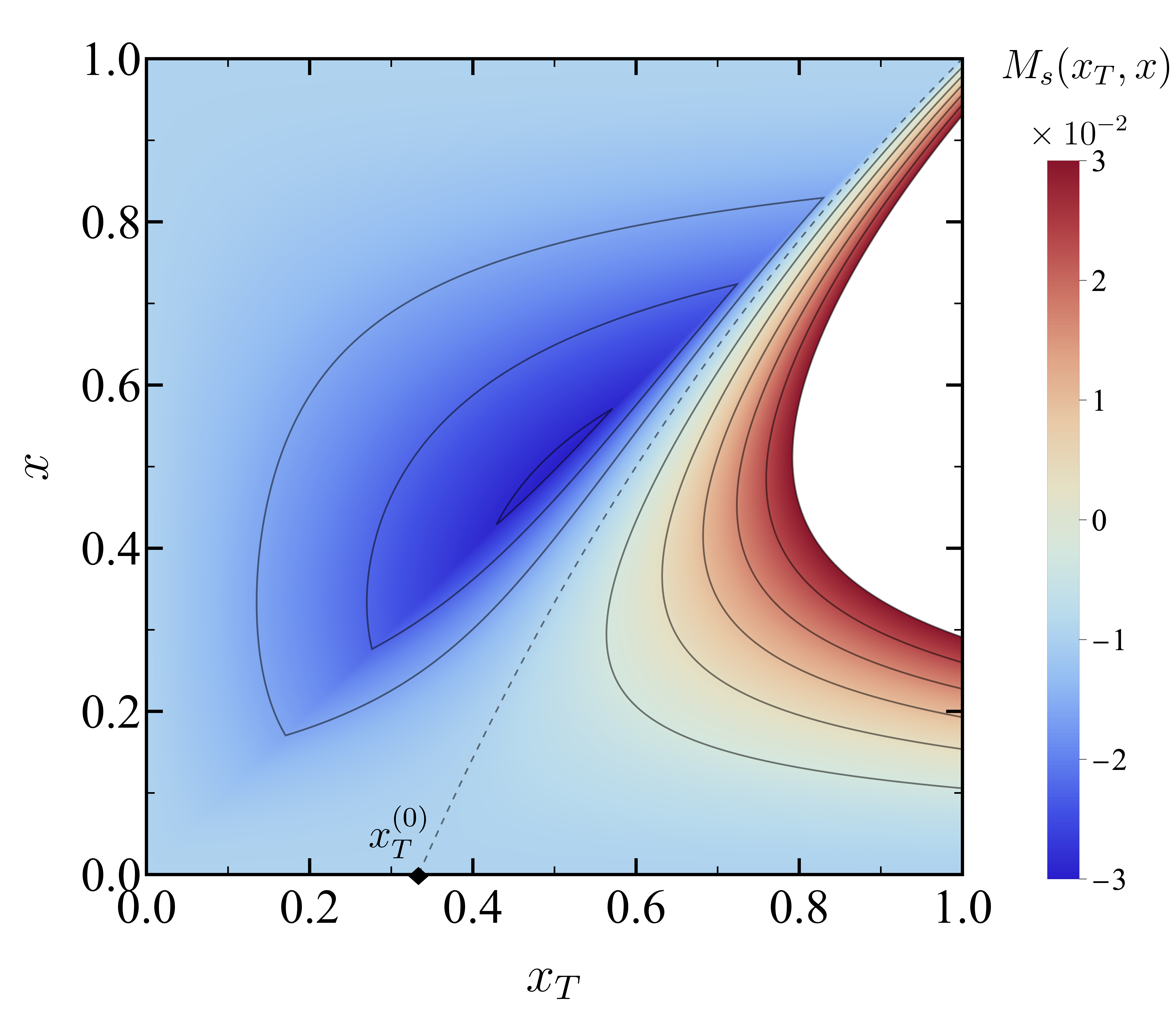}
		\caption{Stability diagram of the nonresetting strategy for heterogeneous resetting. Positive (negative) values, represented as warm (cold) colors, thereof contribute to zero (nonzero) bulk resetting as the best search strategy. White color indicates saturated values. Top: Functional derivative $M(x_T,x)$ defined in Eq.~\eqref{eq:R}. Bottom: Symmetrized functional derivative $M_s(x_T,x)$ defined in Eq.~\eqref{eq:Ms} for the heterogeneous case. The contour line at which $M_s=0$ is marked (dashed line), as well as the value $x_T^{(0)}$ (black diamond) below which $M_s(x_T,x)<0$, $\forall x$.}
		\label{fig:stability-het}
	\end{figure}
	
	We now move to the general case with heterogeneous resetting $r(x)$. Similarly to the homogeneous case, whether or not bulk resetting is advantageous is controlled by the sign of the derivative of the average MFPT with respect to the resetting rate, evaluated at $r(x) \equiv 0$. 
	For heterogeneous resetting, the key quantity is thus
	
	\begin{equation}
		\label{eq:m-het1}
		\mu(x) \equiv \left. \frac{\delta \overline{\tau}}{\delta r(x)}\right|_{r(x)\equiv 0} =\expval{M(x_T,x)}_T,
	\end{equation}
	where the functional derivative $M(x_T,x)$ at fixed target is defined by
	\begin{align}
		M(x_T,x) &\equiv \left. \frac{\delta\tau_0(x_T)}{\delta r(x)}\right|_{r(x)\equiv 0} \nonumber \\ &=\lim_{\varepsilon \to 0^+} \frac{\left.\tau_0(x_T)\right|_{r(x_0)=\varepsilon\delta(x_0-x)}-\left.\tau_0(x_T)\right|_{r(x_0)=0}}{\varepsilon},  
		\label{eq:m-het2}
	\end{align}
	which takes the place of the standard derivative $M(x_T)$ in Eq.~\eqref{eq:m-def}. The physical meaning of $M(x_T,x)$ is clear: It measures how the MFPT varies when the nonresetting strategy is locally perturbed at point $x$. 
	
	The sign of $\mu (x)$ provides a transparent criterion to elucidate the optimality of the nonresetting strategy in the bulk. 
	On the one hand, the positivity of $\mu(x)$ for all $x$ entails that the average MFPT attains a local minimum for $r(x)=0$, since any perturbation from $r(x)=0$ would lead to a larger average MFPT. On the other hand, the emergence of a subinterval where $\mu(x)$ is negative entails the decrease of the average MFPT for a perturbation such that $r(x)\ne 0$ in that subinterval, implying that $r(x)=0$ does not provide the global minimum of the average MFPT. 
	
	The  derivative in Eq.~\eqref{eq:m-het2} can be computed analytically, as detailed in Appendix~\ref{Ap-het}:
	\begin{align}
		\label{eq:R}
		M(x_{T},x)=&\frac{x_T \, x\left(1-|x|\right)^{2}}{2}\Theta\left(-x_T \, x\right) \nonumber \\
		&+\frac{x\left(x_{T}-x\right)\left(1+|x|\right)}{2}\Theta\left(x_T \, x\right)\Theta\left(|x_{T}|-|x|\right)\!\!,
	\end{align}
	
	where $\Theta$ stands for Heaviside's step function. The two terms on the right hand side of Eq.~\eqref{eq:R} have opposite signs: The first  is always negative, whereas the second  is always positive.  Bulk resetting at $x$ is:
	\begin{enumerate}
		\item Beneficial if $x_T \, x<0$ (target and particle positions at opposite sides of the box);
		\item Detrimental for $x_T \, x>0$ and $|x_T|>|x|$ (particle between the center and the target);
		\item Irrelevant if $x_T \, x>0$ and  $|x_T|<|x|$ and thus $M(x_T,x)=0$.
	\end{enumerate}
	
	In the top panel of Fig.~\ref{fig:stability-het}, the function $M(x_T,x)$ in Eq.~\eqref{eq:R} is displayed. For $M(x_T,x) \neq 0$,  the sign of $M(x_T,x)$ is given by the sign of the product $x_T \, x$. Hence, some general statements on the stability of the nonresetting strategy in the bulk can be made without further knowledge of the target distribution. Specifically, $r(x)=0$ does not minimize the MFPT for any target distribution with support in the subintervals $x_T\in (-1,0)$ or $x_T\in (0,1)$. This can be physically understood, because if we know the sign of the target position, the optimal strategy is clearly to forbid exploration of the side of the box opposite to the target's position, i.e., such that $x\, x_T<0$, sending the resetting rate to infinity therein.
	
	We now focus on symmetric target distributions, i.e., $P(x_T) = P(-x_T)$. Therein, the optimal resetting strategy inherits the symmetry of the problem and it suffices to analyze the symmetrized function $M_s(x_T,x)\equiv [M(x_T,x)+M(-x_T,x)]/2$  and its average with respect to $\rhoT(x_T)$ in the interval $[0,1]$. Taking into account Eq.~\eqref{eq:R}, we find the symmetrized function
	\begin{equation}\label{eq:Ms}
		M_s(x_T,x)=
		\begin{cases}
			-\dfrac{1}{4}x^2[1+x+(x-3)x_T], & 0 \leq x \leq x_T ,\\ \, \\
			-\dfrac{1}{4} (1-x)^2x\, x_T, &  x_T \leq x \leq 1.
		\end{cases} 
	\end{equation}
	
	The symmetrized function $M_s(x_T,x)$, displayed in the bottom panel of Fig.~\ref{fig:stability-het}, is thus the key quantity to elucidate whether the strategy with no resetting in the bulk is optimal or not. 
	
	Interestingly, there is an interval of ``small'' target positions, $0\leq x_T\leq x_T^{(0)}$, where $M_s<0$ for all $x$: This means that, for a target distribution with finite support in $(0,x_T^{(0)})$, independently of its detailed shape, $\mu$ is negative for all $x$ and the optimal strategy involves nonvanishing resetting in the bulk regardless of further details of the target distribution. Conversely, there is no interval of ``large'' target positions above which $M_s$ is positive for all $x$ and, thus, we cannot guarantee that a finite-support distribution leads, without further knowledge of its details, to suppression of resetting in the bulk as the optimal strategy. However, looking at the figure, it is clear that the target positions that contribute to $r(x)=0$ as the optimal strategy are those close to the boundaries, i.e., $|x_T|$ close to unity---as already discussed in the simpler case of homogeneous resetting. 
	\begin{figure}
		\centering
		\includegraphics[width = 0.49\textwidth]{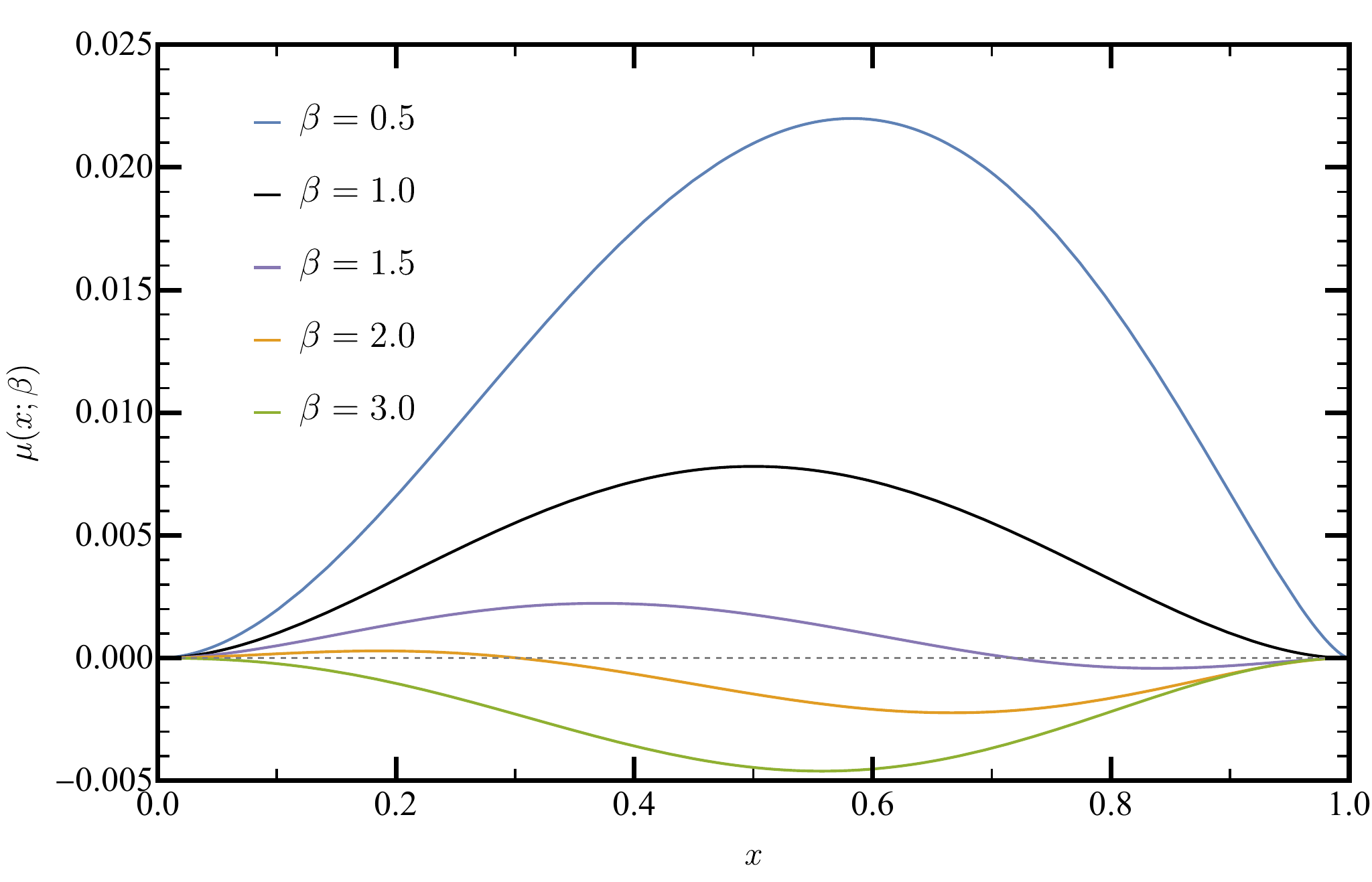}
		\caption{Functional derivative $\mu$ controlling the stability of the nonresetting strategy in the bulk. Specifically, we show $\mu$ as a function of $x$ for the beta family of target distributions introduced in Eq.~\eqref{eq:PTbeta}, given in Eq.~\eqref{SMeq:mu-beta}. The critical value $\beta_c=1$ (black solid line) marks the change to a not-always-positive $\mu(x;\beta)$, which entails an optimal nonzero resetting strategy in the bulk.}
		\label{figSM:het}
	\end{figure}
	
	Now, we turn to studying the optimal resetting strategy for the $\beta$-family of target distributions~\eqref{eq:PTbeta}. Since it is symmetric, $M_s(x_T,x)$ is the function to be integrated in order to obtain $\mu(x;\beta)$---where we have explicitly introduced the dependence on the $\beta$ parameter. 
	In this case, explicit integration of $M_s$ is feasible, with the result
	
	\begin{align}
		\mu(x;\beta)=   \frac{1}{4} x \Bigg\{ &-x (1 + x) + 
		\frac{\Gamma \left(\frac{1}{2} + \beta \right)}{\sqrt{\pi} \, \Gamma(1 + \beta)} \nonumber \\ 
		& \times\Bigg[-(1 - x)^2 + (1 - x)^\beta (1 + x)^{1 + \beta} \nonumber  \\
		&\qquad +2 \beta x^2 (1 + x) \, _2F_1\left(\frac{1}{2}, 1 - \beta; \frac{3}{2}, x^2\right) \Bigg] \Bigg\} \label{SMeq:mu-beta}
	\end{align}
	
	In Fig.~\ref{figSM:het}, we have plotted the functional derivative $\mu$ for different values of $\beta$ as a function of $x$. If for any $x$, $\mu$ takes negative values, this implies that the nonresetting strategy in the bulk is not the optimal one. We have exemplified five values of $\beta$, including the critical value $\beta_c=1$ in black, which stands for the flat distribution. For $\beta \leq 1$, zero bulk resetting is optimal with $\mu (x;\beta) > 0$ for  $x\in (0,1)$. For $\beta>1$, $\mu(x;\beta)$ becomes negative close to the wall $x=1$ (boundary instability), signaling that the nonresetting strategy is no longer the optimal one since $\mu$ becomes negative at least in a certain subinterval.
	Herein, the critical $\beta$ is consistently smaller compared to the homogeneous case, where we had $\beta_c \simeq 1.71$.
	
	\section{Optimal resetting strategy: Numerical results}
	\label{sec-num}
	
	The optimal search strategy with bulk resetting cannot be derived analytically for any $P(x_T)$, because Eq.~\eqref{eqF-compl} does not have an analytical solution for arbitrary $r(x)$. We therefore present numerical solutions, based on gradient descent algorithms~\cite{Folena23physicaa} leading to the optimal $r^*(x)$. The numerical optimization is carried out as follows. The spatial coordinate $x$ is discretized into a mesh of $N$ nodes. An initial search strategy $r_0(x)$ is considered and the average MFPT is computed for such an initial strategy. The following steps are then executed iteratively:
	\begin{enumerate}
		\item For the current search strategy, the functional derivative $\mu$ is computed in the mesh. Details of the calculation are provided in Appendix \ref{ap:num-rec}.
		\item The new potential search strategy is given by $r(x) \to \max[r(x)-\lambda \,\mu,0]$, where $\lambda$ is an adaptive factor with a given value $\lambda_0$ in the first trial.
		\item The average MFPT of the new strategy is computed. If it is lower or equal than the previous one, the strategy is updated and we go to step 1. Otherwise, $\lambda$ is reduced, $\lambda \to \lambda/2$, and we go to step 2.
	\end{enumerate}
	
	As usual, gradient descent is expected to converge to the local minimum whose attraction basin includes the initial condition, and does not provide any information about global optimality. This issue is related to the nonconvexity of the cost function, which can feature multiple local minima. We mitigate this risk by varying the initial resetting profile, $r_0(x)$, either zero or positive uniform values. Numerical parameters typically adopted are $N\in\{501,1001,2001\}$, $\lambda_0\in\{10^5,10^6,10^7\}$.
	The most consuming task in our gradient descent algorithm is to compute the functional derivative, which is detailed in  Appendix \ref{ap:num-rec}.
	
\begin{figure}
		\centering
		\includegraphics[width = 0.49\textwidth]{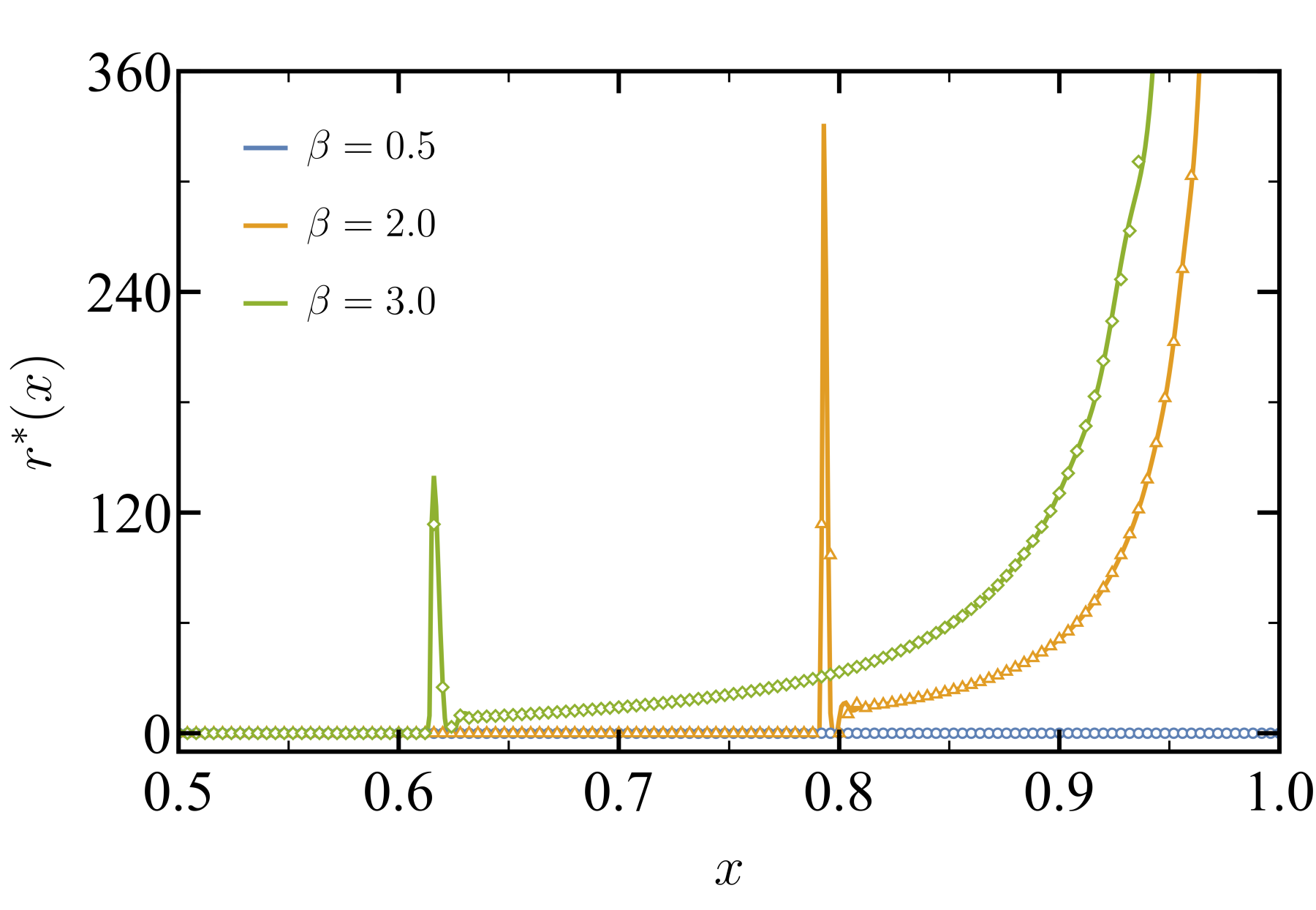}
		\caption{Optimal heterogeneous resetting profiles in the bulk. Specifically, they have been obtained numerically for the target distributions of the family~\eqref{eq:PTbeta} corresponding to $\beta=0.5,2,3$. For $\beta\leq1$, the optimal strategy consistently involves no resetting in the bulk. For $\beta>1$, for which nonzero bulk resetting is the optimal strategy, the features displayed---central region with no resetting for $|x|<x_J$, strongly heterogeneous behavior close to the wall---are robust.  The acceleration factor with respect a vanishing resetting bulk is around 2\% and 7\% for $\beta=2$ and $\beta=3$ respectively. Symbols and lines respectively stand for meshes with $N=501$ and $N=2001$ nodes.}
		\label{fig:optimal_num}
	\end{figure}

	In the case of the beta family of target distributions introduced in Eq.~\eqref{eq:PTbeta}, the optimal strategy obtained displays robust features  as shown in Fig.~\ref{fig:optimal_num}. For $\beta \leq 1$, the numerically optimal resetting strategy converges to vanishing bulk resetting rate as theoretically predicted. For $\beta>1$, there appears a value $x_J$ that separates zero resetting and nonzero resetting regions, i.e., $r^*(x)=0$ for $0<|x|<x_J$,  whereas $r^*(x)\ne 0$ for $x_J<|x|<1$. Resetting is suppressed in $|x|<x_J$  to ensure uninterrupted exploration. For $|x|>x_J$, resetting is significantly heterogeneous, displaying quite high values close to the walls, i.e., for $|x|\to1$. At $|x|=x_J$ the numerical results evidence the presence of a Dirac delta contribution to $r(x)$, which is corroborated by the use of different meshes---the peak amplitude is proportional to $N$ with a mesh size scaling as $1/N$. Interestingly, we observe that the qualitative features of the profile for the optimal resetting obtained, i.e., the vanishing window, the peak and the high heterogeneity, are robust for other families with accumulation of probability in the center, as detailed in Appendix~\ref{ap:rob}. 
	
	One of the most remarkable features of our optimal resetting profile is the emergence of Dirac-delta contributions. To the best of our knowledge, this is the first instance of such a resetting strategy. Although peculiar, delta-like resetting for certain points has no physical issues: It can be regarded, for instance, as a limit process in a narrow spatial window, inside which one has a large homogeneous resetting rate. As this spatial window becomes narrower, the large homogeneous resetting rate increases while the area remains constant. A heuristic interpretation of a Dirac-delta resetting and its practical implementation are given in Appendix~\ref{ap:dirac}. For some target distributions, the optimal resetting strategy consists of two symmetric delta peaks only. Specifically, let us consider the family
	\begin{equation}
		\label{SMeq:poly}
		P(x_T;a)=a x_T^2 + (4 - 3 a) x_T^4 + \frac{7}{30} (-9 + 8 a) x_T^6,
	\end{equation}
	with $0<a<(117 + \sqrt{8169})/23$. These specific coefficients are chosen to have a monoparametric distribution family, and the interval of $a$ stems from $p(x_T;a)\geq 0$ for all $0<x_T<1$. When the optimal resetting is different from the nonresetting strategy in the bulk, the optimal strategy comprises two symmetric Dirac delta functions at specific positions and with specific intensity that can be numerically derived. 
	\begin{figure}
		\centering
		\includegraphics[width = 0.49\textwidth]{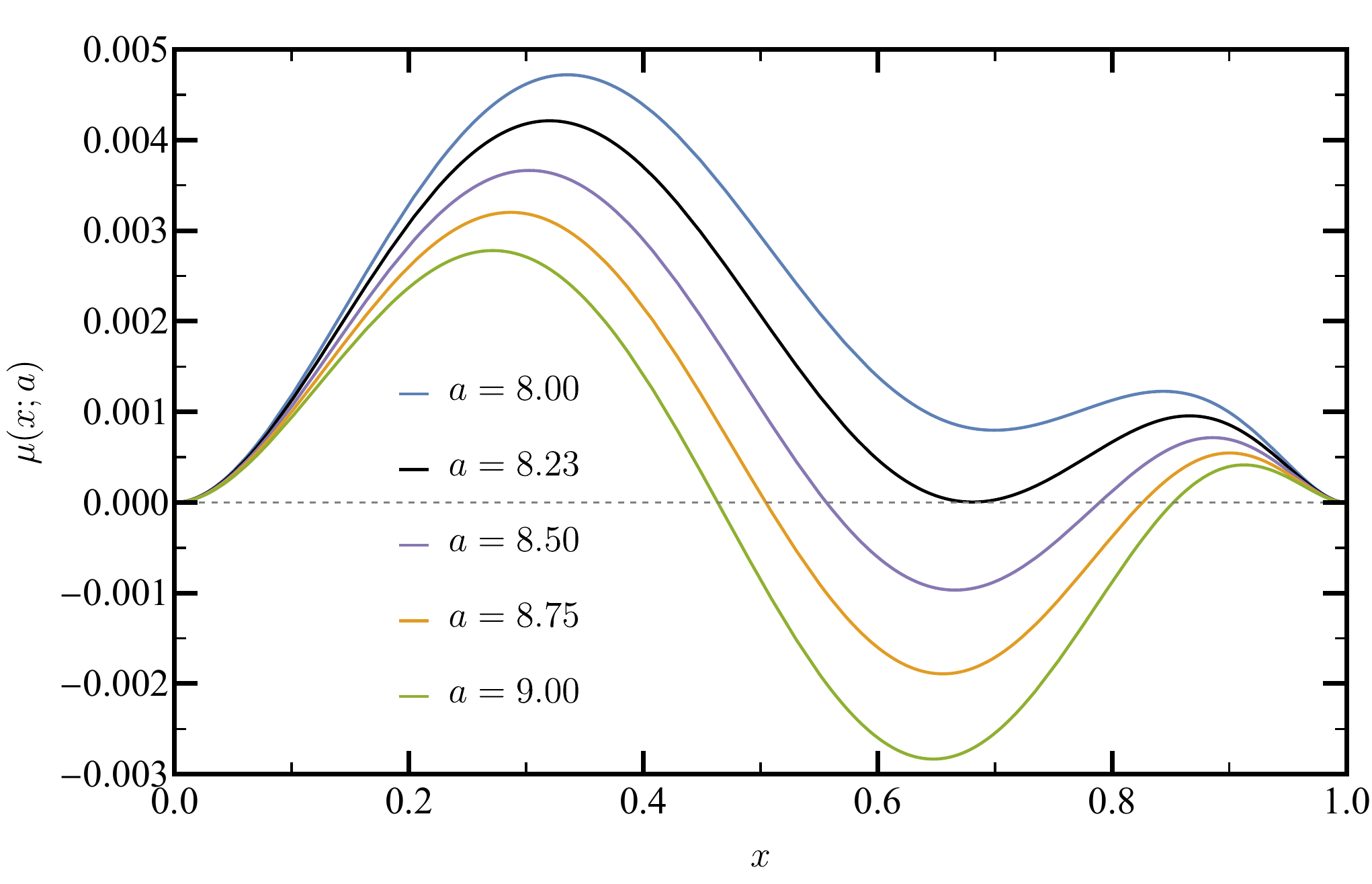}
		\caption{Functional derivative  controlling the stability of the nonresetting strategy for the polynomial family in Eq.~\eqref{SMeq:poly}. The critical value $a_c$ (black solid line) marks the change where $\mu(x;a)$ is not always positive, indicating an optimal nonzero resetting strategy.}
		\label{figSM:poly0}
	\end{figure}
	We prove first that the optimal resetting strategy displays a transition for the family of target distributions in Eq.~\eqref{SMeq:poly} when varying the control parameter $a$. To do that, we have analytically computed $\mu(x;a)$ for this target distribution. The result is a tenth-degree polynomial that is strictly positive in the interval $0<x<1$ for $a<a_c\simeq 8.23$ as shown in Fig.~\ref{figSM:poly0}. In this case, $\mu(x;a)$ for $a>a_c$ becomes negative in at least a subinterval of the bulk, $x\in[0,1]$; we call this behavior a bulk instability, as opposed to the boundary instability defining the optimality criteria for the beta distribution. 
	
	For $a>a_c$, we address a constrained optimization problem for the average MFPT over the family of strategies given by
	\begin{equation}
		r_\delta (x) = \frac{u}{x_R} [\delta(x-x_R)+\delta(x+x_R)].\label{eq:ProfileResetting_Poly}  
	\end{equation}
	The computation of the average MFPT can be analytically performed, obtaining $\overline{\tau}$ as a function of $x_R$ and $u$. Once this result is obtained, numerical optimization can be carried out over $x_R$ and $u$. We note that this constrained optimization does not guarantee that Eq.~\eqref{eq:ProfileResetting_Poly} is the optimal strategy, but the analytical computation of the functional derivative for this resetting profile proves that it gives a (local) minimum of the MFPT: The resulting functional derivative is nonnegative, being equal to zero at the boundaries and at $x=\pm x_R$, proving that the heterogeneous resetting strategy in Eq.~\eqref{eq:ProfileResetting_Poly} is locally optimal. 
	\begin{figure}
		\begin{center}
			
			\includegraphics[width=\linewidth]{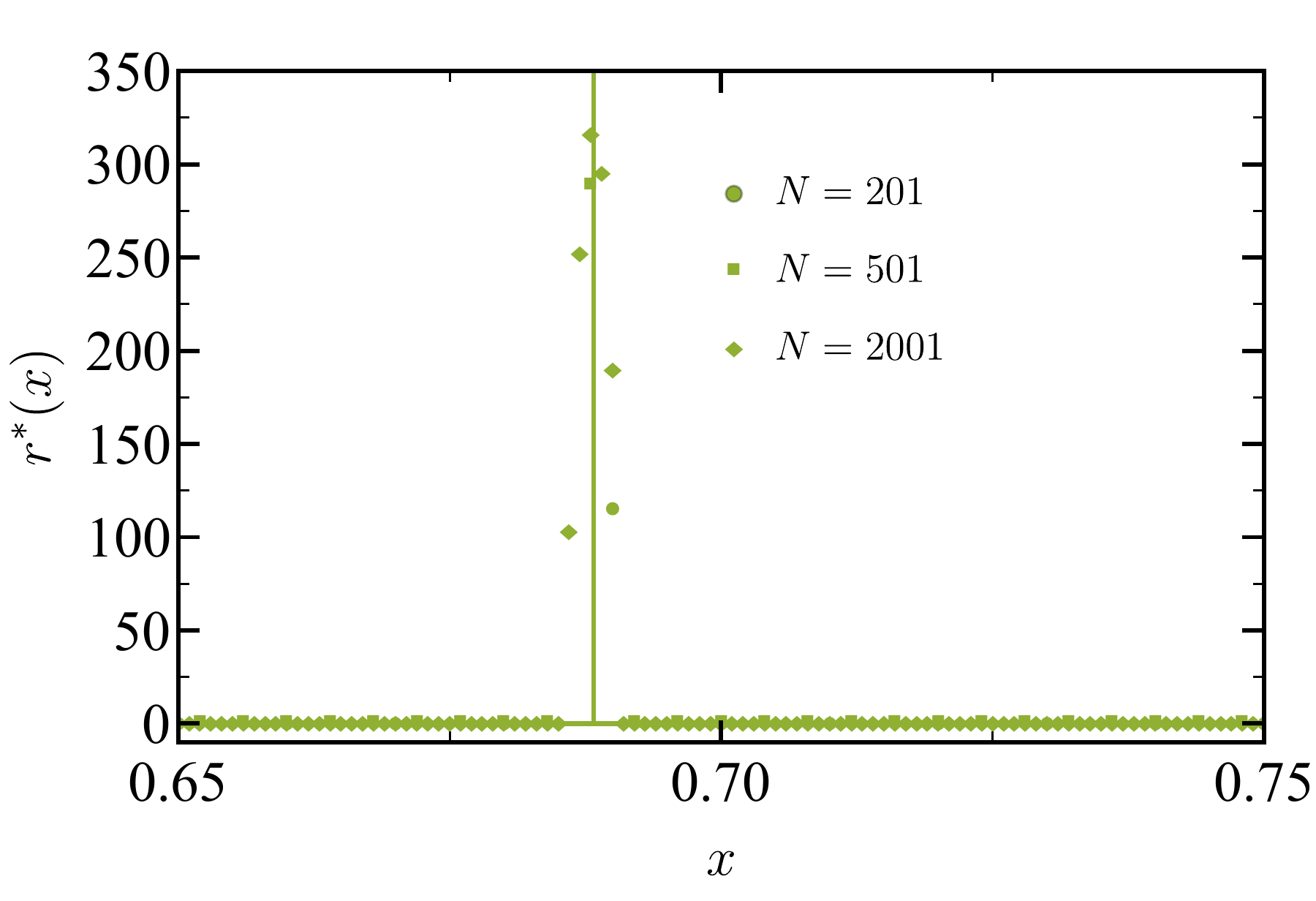}
			
			\hfill
			
			\centering
			\includegraphics[width=\linewidth]{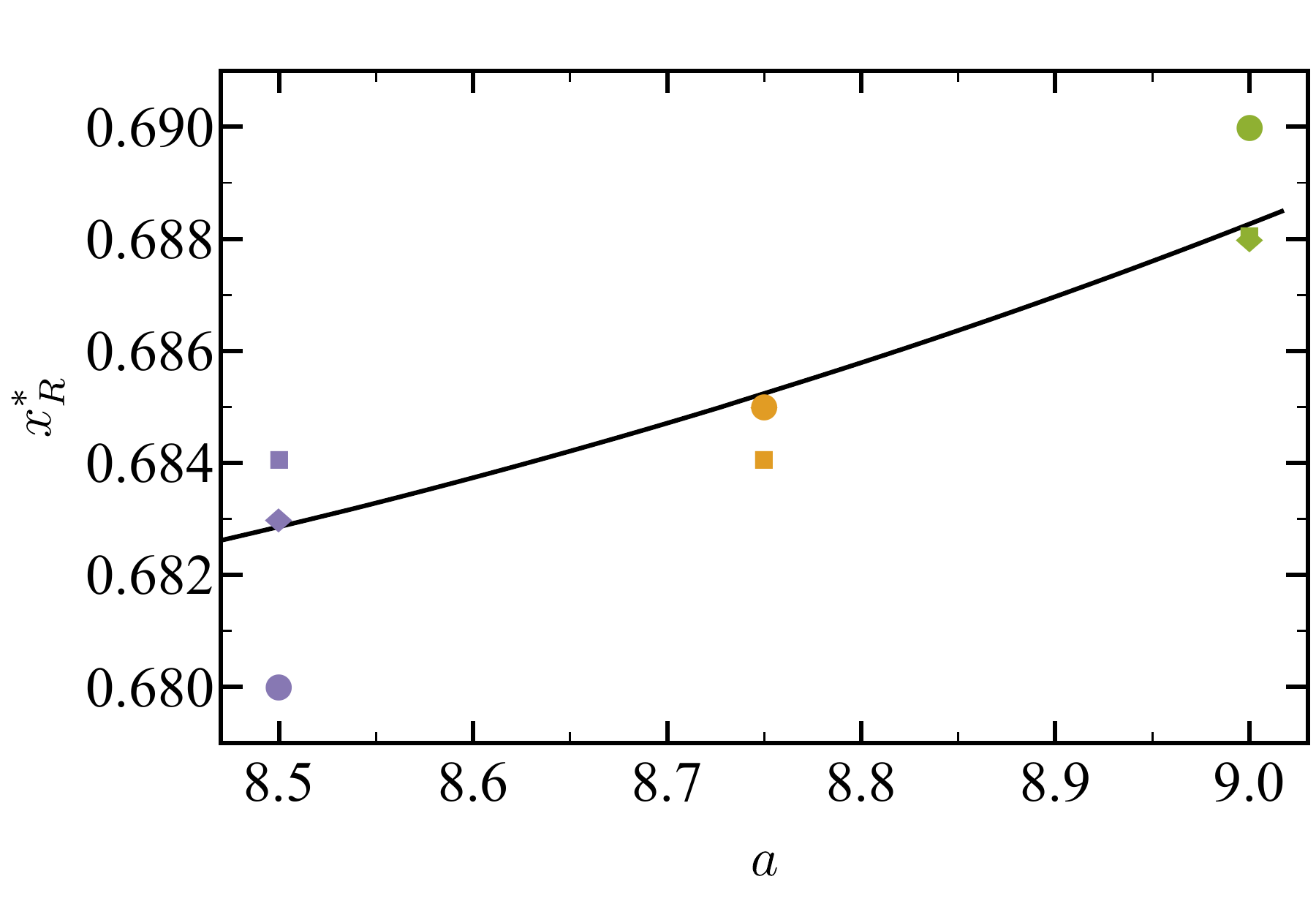}
			
			\centering
			\includegraphics[width=\linewidth]{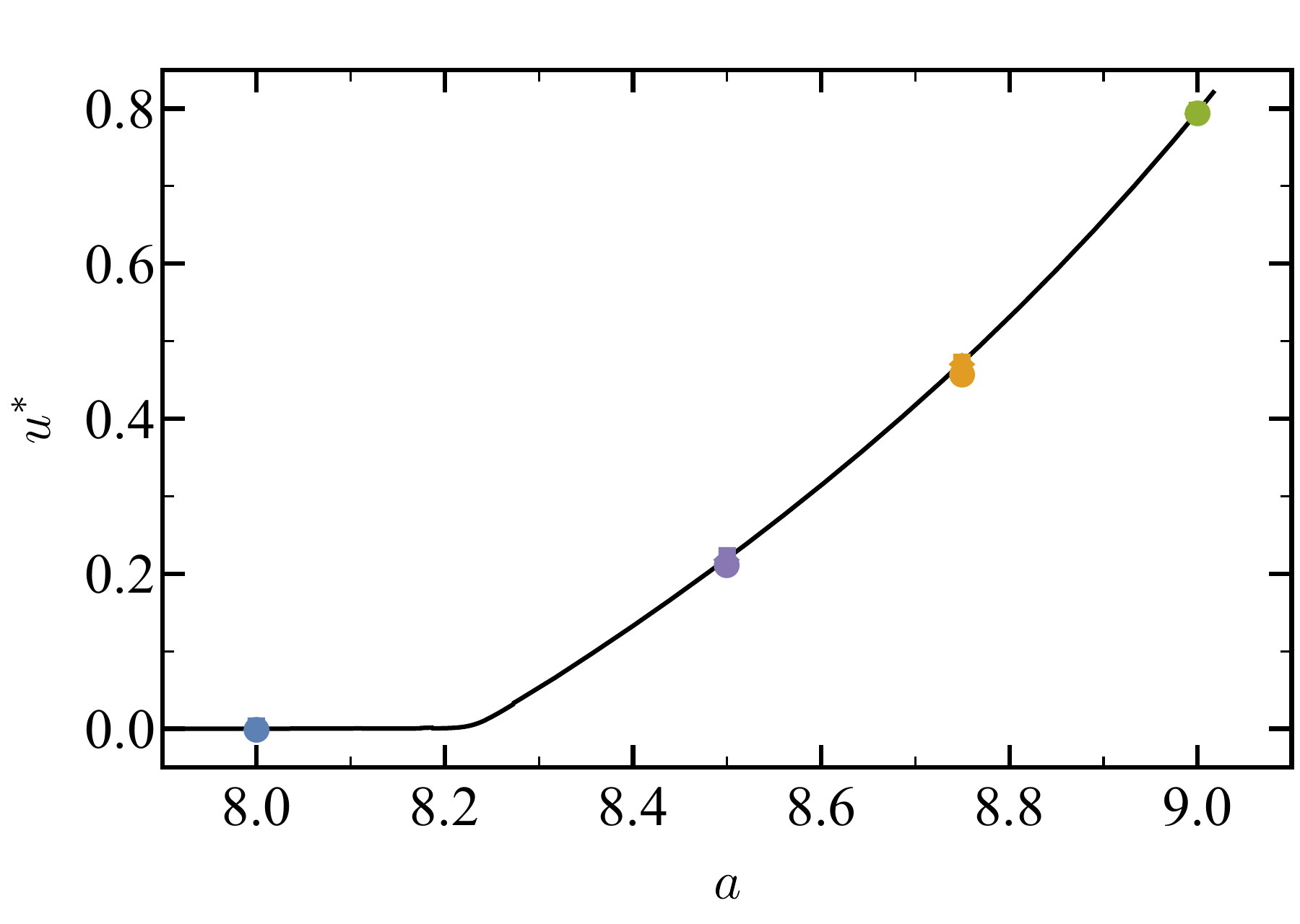}
			
		\end{center}
		\caption{Top: Optimal strategy for the polynomial family in Eq.~\eqref{SMeq:poly}. The solid lines correspond to the theoretical optimal strategy, Dirac deltas characterized by $x_R^*(a)$ and $u^*(a)$, whereas the symbols correspond to the numerically obtained optimal solution for different grid meshes.  Center and Bottom: Optimal parameters for the resetting strategy. The theoretical parameters (black solid lines) are compared to the numerical parameters obtained by gradient descent optimization using $N=201, \, 501, \, 2001$ (circles, squares and diamonds, respectively).}
		\label{figSM:poly0op}
	\end{figure}
	Furthermore, we have run numerical optimization for several values of $a$, and we have observed the optimal profile to be consistent with our theoretical expectation. For a dense mesh, we obtain a sharply peaked resetting strategy, as shown in the top panel of  Fig.~\ref{figSM:poly0op}. In the central and bottom panels, we show that the values of the position of the Dirac-delta and its intensity, measured in the numerical optimization through its area, are consistent---error bars are willingly eluded to improve the visualization but, when taken into account, numerical results are compatible with the theoretical expectations.
	
	\section{Conclusions}
	\label{sec-concl}
	
	In this study, we have obtained the optimal resetting strategies for search processes in a confined $1d$ geometry with distributed targets. We have introduced resetting boundaries, which are always beneficial for our goal: If the searcher has not found the target upon reaching the wall, it is advantageous to be reset to the initial point. We have then analyzed the impact of the resetting strategy in the bulk, taking into account the role of the spatial distribution of the target. We have provided rigorous mathematical criteria to determine when the absence of resetting in the bulk is optimal, under both homogeneous and heterogeneous resetting rates. For homogeneous resetting, the convexity of the relevant function guarantees that, if the MFPT has a local minimum for $r=0$, this is indeed the global minimum. For heterogeneous resetting, our approach rigorously determines whether $r(x)=0$ constitutes a local minimum, since no general convexity argument holds in this case. Remarkably, these criteria, valid for arbitrary target distributions, have been further illustrated by considering two monoparametric families of such distributions---allowing us to shed further light on the physical reasons behind the optimality of bulk resetting.
	
	Our work paves new research lines in the study of search processes. On the one hand, we have shown that the spatial distribution of the target plays a key role in determining the optimal search strategy. This contrasts with the typical approach of optimizing the search strategy for a fixed, known position of the target: In an actual search process, the searcher does not know the position of the target. On the other hand, the optimal resetting strategy is highly nontrivial and heterogeneous---including Dirac-delta peaks that have not been previously reported, even for the relatively ``simple'' one-dimensional case analyzed in this work. Several future research directions arise: (i) to better understand the physical origin of Delta-like contributions in the optimal resetting rate; (ii) to derive analytical expressions for it; (iii) to extend our analysis to distributions with support over the full real axis and (iv) to searches in multidimensional domains.
	
	\begin{acknowledgements}
		
		G.~García-Valladares, C.~A.~Plata and A.~Prados acknowledge financial support from Grant PID2024-155268NB-I00 funded by MICIU/AEI/10.13039/501100011033/ FEDER, UE and Grant ProyExcel\_00796 funded by Junta de Andalucía's PAIDI 2020 programme. C.~A.~Plata and A.~Prados acknowledge financial support from the applied research and innovation Project PPIT2024-31833, cofunded by EU--Ministerio de Hacienda y Función Pública--Fondos Europeos--Junta de Andalucía--Consejería de Universidad, Investigación e Innovación. A.~Manacorda acknowledges financial support from the project “MOCA” funded by MUR PRIN2022 Grant No.~2022HNW5YL.
		
	\end{acknowledgements}
	
	\section*{Data availability}
	
	Numerical codes employed for generating the data and figures that support the findings of this study are openly available in the GitHub page of University of Sevilla's FINE research group~\cite{github-optimal-resetting}.
	
	\appendix
	
	\section{Exact calculations for homogeneous resetting}
	\label{Ap-hom}
	
	The average MFPT can be computed by integrating Eq.~\eqref{eq:sol-r-const}  along with the target distribution. Since we focus on the stability of the nonresetting strategy in the bulk, it is useful to carry out a Taylor expansion around $r=0$, which for the case of resetting boundaries yields 
	\begin{equation}
		\overline{\tau}=\frac{\left\langle x_T^2 + \left|x_T\right|\right\rangle_T}{2}+ \frac{\left\langle x_T^4 - \left|x_T\right| \left( 1-2x_T^2\right) \right\rangle_T}{24} r + \mathcal{O}(r^2).
		\label{SMeq:hom_exp}
	\end{equation}
	The linear term in $r$ is the key quantity that has been introduced in Eq.~\eqref{eq:m-def}.
	
	For the family of target distributions introduced in Eq.~\eqref{eq:PTbeta}, explicit integration of the average MFPT is feasible, resulting in
	\begin{align}
		\overline{\tau}(r;\beta) = &r^{-1} \Bigg[ {}_0F_1 \left(; \beta+\frac{1}{2} ; \frac{r}{4}\right)  -1 +2^{\beta-\frac{1}{2}} r^{\frac{1}{4}-\frac{\beta}{2}}\nonumber \\
		&\times 
		\Gamma\left( \beta +\frac{1}{2}\right) 
		L_{\beta-\frac{1}{2}} \left(\sqrt{r}\right) \tanh \left( \frac{\sqrt{r}}{2}\right) \Bigg],
		\label{SMeq:hombeta}
	\end{align}
	where ${}_0F_1(;b;z)$ stands for the confluent hypergeometric function and $L_{\alpha}(x)$ for the modified Struve function.
	
\begin{figure}
		\begin{center}
			\includegraphics[width=0.49\textwidth]{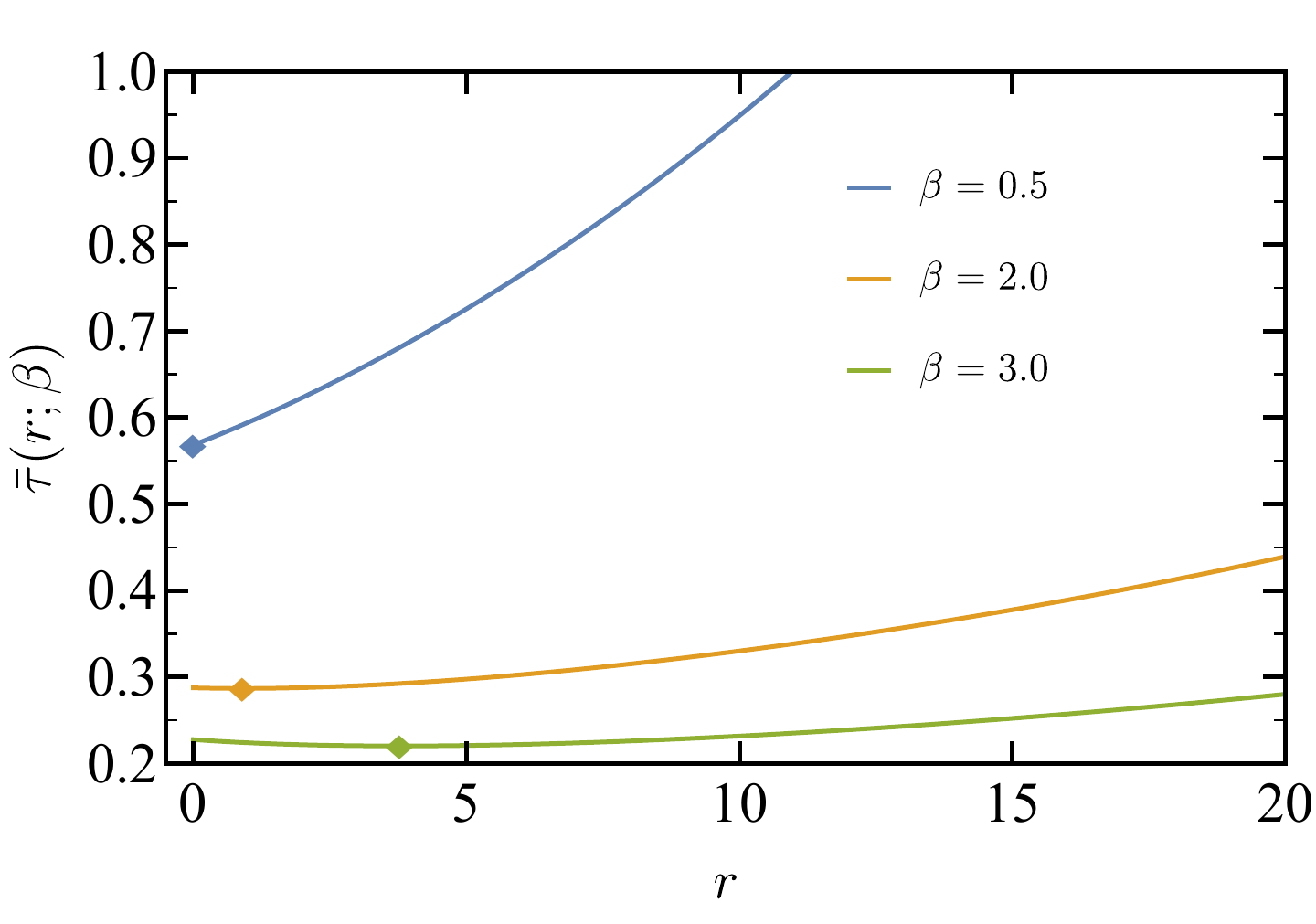}
			\includegraphics[width=0.49\textwidth]{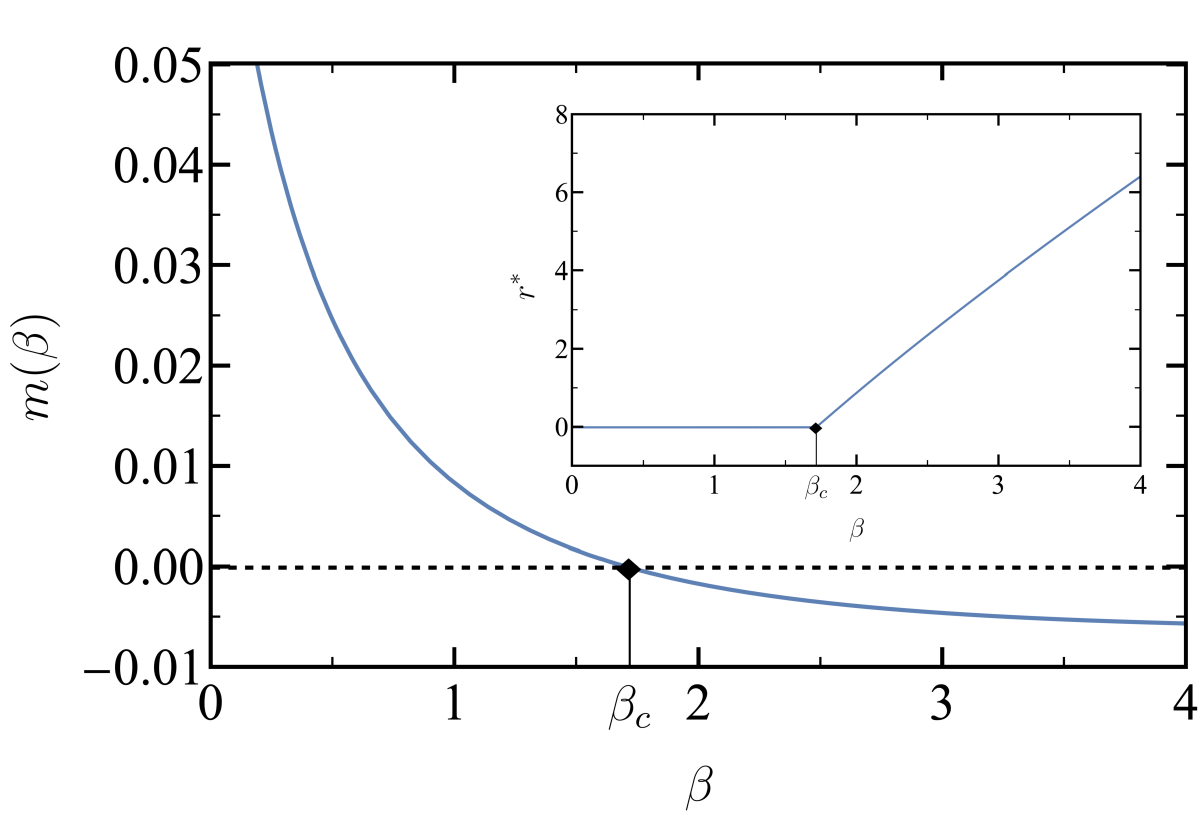}
		\end{center}
		\caption{Top: Average MFPT for different beta distributions, as given by Eq.~\eqref{SMeq:hombeta}. For $\beta=0.5$, the minimum is attained at $r^*=0$ (blue diamond) while for $\beta=2,\, 3$ is attained at $r^*>0$ (orange and green diamonds, respectively). The acceleration factor for the optimal value with respect a vanishing resetting bulk is around 0.2\% and 3\% for $\beta=2$ and $\beta=3$ respectively Bottom: Slope of the average MFPT at $r=0$, as shown in Eq.~\eqref{eq:m-hom-beta} of the main text. It crosses zero just at $\beta=\beta_c\simeq 1.71$ (black diamond). In the inset, the continuous phase transition for the optimal homogeneous rate $r^*$ is shown.}
		\label{figSM:hom}
	\end{figure}

	In Fig.~\ref{figSM:hom}, we have displayed  the average MFPT for different values of $\beta$ as function of $r$ in the top panel. As discussed in the main text, the minimum is achieved either at $r^*=0$ or at $r=r^*>0$ if $\beta$ is lower or greater than the critical value $\beta_c$, respectively. This critical value is the only one that makes the function $m(\beta)$ reported in Eq.~\eqref{eq:m-hom-beta} vanish,  and it is marked in the bottom panel with a black diamond. Optimal values of the homogeneous resetting rate are visualized in the inset as a function of $\beta$, displaying the usual behavior of a continuous phase transition. 
	
	\section{Exact calculations for heterogeneous resetting}
	\label{Ap-het}
	
	We focus on the stability of $r(x)=0$ as the optimal strategy. Hence, we study the expansion of the average MFPT, which is now a functional expansion, up to linear order in $r(x)$: 
	\begin{align}
		\overline{\tau}[r]= &\frac{\langle x_T^2 + |x_T| \rangle_T}{2} \nonumber \\ 
		&+ \int dx \left\langle \left.  \frac{\delta\tau_0(x_T)}{\delta r(x)}\right|_{r=0}\right\rangle_T r(x) + \mathcal{O}((r(x))^2).
	\end{align} 
	The coefficient of $r(x)$ within the integral is the function $\mu(x)$ introduced in Eq.~\eqref{eq:m-het1}, which is the key quantity for discussing if the nonresetting strategy in the bulk is locally optimal or not. Consistently, the value for $r=0$ is equal to the one obtained in the homogeneous expansion of Eq.~\eqref{SMeq:hom_exp}.
	
	For computing the function $\mu(x)=\langle M(x_T,x)\rangle_T$, we resort to the definition of functional derivative in Eq.~\eqref{eq:m-het2}.
	Therefore, we need to solve the equation for the MFPT in Eq.~\eqref{eqF-compl} with $r(x_0)$ being a Dirac-delta at $x_0=x$. This is possible, since we can solve the equation separately for $x_0<x$ and $x_0>x$ with $r=0$, and enforce (i) the continuity of $F$ at $x_0=x$ and (ii) the kick condition $F'(x^+|x_T)-F'(x^-|x_T)=\varepsilon F(x|x_T)$, $F'$ is discontinuous because of the Dirac delta kick. The final solution of $F(x_0|x_T)$ has to be evaluated at $x_0=x_T$ to obtain the MFPT. After a careful calculation, it is obtained
	
	\begin{align}
		\delta \tau_0\equiv & \tau_0[r(x_0)=\varepsilon \delta(x-x_0)] - \tau_0[r(x_0)=0]\nonumber \\
		=&-\dfrac{\varepsilon}{2} \dfrac{\vert x\, x_T \vert (1-\vert x\vert)^2}{1+\varepsilon \vert x \vert (1-\vert x \vert)}\Theta\left(-x_T \, x\right) \nonumber \\
		&+ \dfrac{\varepsilon}{2} |x|(|x_T|-|x|)(1+|x|)
		\Theta\left(x_T \, x\right)\Theta\left(|x_{T}|-|x|\right)\!\!,
		\label{SMeq:cases1}
	\end{align}
	where $\tau_0[r(x_0)=0]=\frac{1}{2}(x_T^2+|x_T|)$, consistent with our previous results in the absence of resetting, and $\Theta$ stands for Heaviside's step function. Note that for $\varepsilon>0$, $\delta \tau_0$ is either negative for $x_T \, x<0$, positive for $x_T \, x>0$ and $|x_T|>|x|$  or 0 for $x_T \, x>0$ and $|x_T|<|x|$, as physically expected. Substituting Eq.~\eqref{SMeq:cases1} into Eq.~\eqref{eq:m-het2} and taking the limit $\varepsilon\to 0^+$, we get Eq.~\eqref{eq:R}.
	
	For symmetric target distributions, it suffices to focus on the symmetrized function $M_s(x_T,x)=[M(x_T,x)+M(-x_T,x)]/2$---defined for $(x_T,x) \in [0,1]^2$, which results in Eq.~\eqref{eq:Ms}.
	This symmetrized function, which facilitates our general discussion on the stability of the nonresetting strategy in the bulk in the main text, has been displayed in the bottom panel of Fig.~\ref{fig:stability-het}. In that plot, we have highlighted the contour at which $M_s=0$, 
	\begin{equation}
		\left. x(x_T)\right|_{M_s=0}= \frac{3x_T-1}{1+x_T}.
	\end{equation}
	This contour hits the horizontal axis ($x=0$) at $x_T^{(0)}=1/3$. 
	
	The form of the symmetric cost $M_s(x_T,x)$ gives us direct information on the cost derivative
	\begin{equation}
		\mu(x) = 2 \int^1_0 \mathrm{d}x_T \, P(x_T) M_s(x_T,x) \ .
	\end{equation}
	Equation~\eqref{eq:R} directly shows that $\mu(x=0) = \mu(x=\pm1) = 0$. In the symmetric case, it is easy to see from Eq.~\eqref{eq:Ms} that $\mu'(x=0) = \mu'(x=1) = 0$. The behavior close to the boundaries is thus given by the second derivative, which reads
	\begin{subequations}
		\label{SMeq:ddmu}
		\begin{align}
			\left. \mu''(x) \right\vert_{x=0} &= \frac{3}{2} \left\langle \vert x_T \vert \right\rangle_T -\frac{1}{2} \\ \left. \mu''(x) \right\vert_{x=1} &= - \frac{1}{2}\left\langle \vert x_T \vert \right\rangle_T +  P(1) \ .
		\end{align}
	\end{subequations}
	Equations~\eqref{SMeq:ddmu} provide analytical criteria for the optimality of zero bulk resetting. Convexity at $x=0,1$ is a \textit{necessary} condition for optimality,  i.e., $\mu''(x=0,1)>0$. This leads to the simple criterion 
	\begin{equation}\label{SMeq:crit}
		1/3 < \left\langle \vert x_T \vert \right\rangle_T < 2 P(1) \ ,
	\end{equation}
	which has a clear physical interpretation: Unless the target is close enough to the origin or the target distribution decays significantly at $x=1$, resetting should be avoided. 
	
	It must be emphasized that Eq.~\eqref{SMeq:crit} guarantees the stability of nonresetting in the bulk at the boundaries of the interval $x \in [0,1]$, but instabilities can also appear inside the interval, e.g., if $P(x_T)$ becomes small at intermediate $x_T$, see the discussion about the family of target distributions introduced in Eq.~\eqref{SMeq:poly}. Summing up, Eq.~\eqref{SMeq:crit} is a neccessary---but not sufficient---condition for having a locally minimum MFPT for $r(x)=0$.
	
	\section{Numerical recipe for the optimization problem}
	\label{ap:num-rec}
	
	This Appendix is devoted to the computation of the variation $M(x_T,x)$, defined for the variation with respect to $r(x)=0$ in~\eqref{eq:m-het2}, but with respect an arbitrary resetting profile. We are interested in computing the general functional derivative, 
	\begin{align}
		M(x_T,x) \equiv & \frac{\delta \tau_0(x_T)}{\delta r(x)} \nonumber \\ 
		=& \lim_{\varepsilon \to 0 } \varepsilon^{-1} \left\{ \tau_0(x_T)[r(x_0)+\varepsilon \delta(x_0-x)] \right. \nonumber\\
		&\qquad \qquad \left.-\tau_0(x_T)[r(x_0)]\right\},
	\end{align}
	where we stress in the displayed notation that $\tau_0(x_T)[r(x_0)]$ is the MFPT to the target located at $x_T$ of our searcher with a resetting strategy given by the resetting profile $r(x_0)$. This is a generalization of Eq.~\eqref{eq:m-het2}, where the perturbation is made around $r(x_0)=0$, instead of around an arbitrary $r(x_0)$. 
	
	The MFPT is obtained as $\tau_0(x_T)=-F(x_T)$, where $F$ satisfies Eq.~\eqref{SM:eqF}. 
	
	For building the solution at $x_0=x_T$, it suffices to solve the equation in the interval with boundaries $x_T$ and $-\sigma(x_T)$
	
	. This means that it suffices to consider the boundary conditions
	\begin{align}
		F(-\sigma (x_T))=0, \quad 
		F(0)=0.
	\end{align}
	If we consider the perturbed solution, i.e., $r(x_0) \to r(x_0)+ \delta(x_0-x)$, we can consider the same equation but with the extra matching conditions
	\begin{align}
		F(x^+)-F(x^-)&=0,\\
		F'(x^+)-F'(x^-)&=\varepsilon F(x).
	\end{align}
	We denote the solution of the previous equations by $\tilde{F}(x_T,x)$, evaluated at $x_0=x_T$ when considering the perturbation at $x_0=x$.  Note that
	\begin{equation}
		M(x_T,x)=- \left. \frac{d\tilde{F}(x_T,x)}{d\varepsilon}\right|_{\varepsilon=0}.
	\end{equation}
	The difference $\Delta F=F-\left. F\right|_{\varepsilon=0}$ between the perturbed and the unperturbed solutions fulfills
	\begin{align}
		\Delta F '' (x_0)&= r(x_0) \Delta F(x_0), \\
		\Delta F (-\sigma(x_T))&=0,\\
		\Delta F (0) &=0,\\
		\Delta F(x^+)- \Delta F(x^-) &=0,\\
		\Delta F'(x^+)- \Delta F'(x^-) &= \varepsilon F(x). 
	\end{align}
	It is obvious from the definition of $\Delta F$ that we have also
	\begin{equation}
		M(x_T,x)=- \left. \frac{d \Delta \tilde{F}(x_T,x)}{d\varepsilon}\right|_{\varepsilon=0}.
	\end{equation}
	
	\begin{widetext}	
		It will be handy to use the ``fundamental'' solutions $F_1$, $F_2$, $F_3$, and $F_4$, defined by 
		\begin{align}
			F_1''(x_0) &= r(x_0) F_1(x_0) - 1 \ , & F_2''(x_0) &= r(x_0) F_2(x_0) \ , & F_3''(x_0) &= r(x_0) F_3(x_0) - 1 \ , & F_4''(x_0) &= r(x_0) F_4(x_0) \ , \\
			F_1(-\sigma(x_T)) &= 0 \ , & F_2(-\sigma(x_T)) &= 0 \ , & F_3(0) &= 0 \ , & F_4(0) &= 0 \ , \\
			F'_1(-\sigma(x_T)) &= 0 \ , & F'_2(-\sigma(x_T)) &= 1 \ , & F'_3(0) &= 0 \ , & F'_4(0) &= 1 \ , 
		\end{align}
		for building the different solutions. 
	\end{widetext}
	
	It is convenient to write some of our second-order differential equations as a system of two first-order differential equations. By defining $G=F'$, our homogeneous equations can be written as
	\begin{equation}
		\left(\begin{array}{c}
			F'(x_{0})\\
			G'(x_{0})
		\end{array}\right)=\underbrace{\left(\begin{array}{cc}
				0 & 1\\
				r(x_{0}) & 0
			\end{array}\right)}_{A(x_{0})} \underbrace{\left(\begin{array}{c}
				F(x_0)\\
				G(x_0)
			\end{array}\right)}_{\vec{v}(x_0)} , 
	\end{equation}
	or, in  matrix form, as
	\begin{equation}
		\frac{d}{dx_0} \vec{v}(x_0) = A(x_0) \vec{v}(x_0).
	\end{equation}
	If one has initial conditions for $x_0=x_i$, i.e., $\vec{v}(x_i)=\vec{v}_i$, the solution can be  built with the propagator $U(x_0,x_i)$, 
	\begin{equation}
		\vec{v}(x_0) = U(x_0,x_i) \vec{v_i},
	\end{equation}
	where the propagator is an operator fulfilling  
	\begin{equation}
		\frac{\partial}{\partial x_0} U(x_0,x_i) = A(x_0) U(x_0,x_i) 
	\end{equation}
	with the initial condition $U(x_i,x_i)=I$, being $I$ the identity operator. 
	
	In the following, we study in detail the case $x_T>0$. This choice makes $\sigma(x_T)=+1$. For negative $x_T$, we can use that $\tilde{F}(-x_T,-x)=\tilde{F}(x_T,x)$.  We split the study of $M(x_T,x)$ into different cases, depending on the relative positions between $x$ and $x_T$. As already stated, we focus on the computation for $x_T>0$.
	
	\textbf{Case I:  $x>0$, $x>x_T$} (general case: $x\, x_T>0$, $|x|>|x_T|$)
	
	This is the easiest situation, there is no difference between $\tilde{F}(x_T,x$) and $\left. \tilde{F}(x_T,x) \right|_{\varepsilon=0}$ since $x>x_T$. Consequently, $M(x_T,x)=0$.
	
	\textbf{Case II:  $x>0$, $x<x_T$} (general case: $x\, x_T>0$, $|x|<|x_T|$)
	
	In this case, one needs to solve the equation in the interval $x_0 \in (-1,x)$, which can be done using the conditions at $x_0=-1$ and $x_0=0$. At $x_0=x$, $F$ is continuous but $F'$ experiments a kick stemming from the Dirac-delta perturbation, and the solution can be propagated from $x^+$ up to $x_T$. In terms of $F(x)$---we will compute it afterwards, one can write
	\begin{equation}
		\left(\begin{array}{c}
			\Delta F(x_T)\\
			\Delta F'(x_T)
		\end{array}\right) = U(x_T,x) \left(\begin{array}{c}
			0\\
			\varepsilon F(x)
		\end{array}\right).
	\end{equation}
	Hence,
	\begin{equation}
		M(x_T,x)=- \left. \frac{d \Delta F(x_T,x)}{d\varepsilon} \right|_{\varepsilon=0} =U_{12}(x_T,x) F(x).
	\end{equation}
	We have computed the functional derivative for the case $0<x<x_T$ except for the value of $F(x)$, which can be obtained by a convenient combination of the solutions $F_1$ and $F_2$. In the interval $x_0 \in (-1,x)$, we write the solution as
	\begin{equation}
		F(x_0)= F_1(x_0)+ \alpha F_2(x_0).
	\end{equation} 
	Enforcing $F(0)=0$, we get
	\begin{equation}
		\alpha=-\frac{F_1(0)}{F_2(0)}.
	\end{equation}
	So, we have
	\begin{equation}
		F(x_0)= F_1(x_0) -\frac{F_1(0)}{F_2(0)} F_2(x_0),
	\end{equation} 
	which substituted into the previous equation leads us to the final result
	\begin{equation}
		M(x_T,x)=U_{12}(x_T,x) \left[F_1(x) -\frac{F_1(0)}{F_2(0)} F_2(x) \right].
	\end{equation}
	
	\textbf{Case III:  $x<0$} (general case: $x\, x_T<0$)
	
	Here, we need to solve $F$ for both $x_0 \in (-1,x)$ and $x_0 \in (x,x_T)$. This can be done by properly using the fundamental solutions: We take
	\begin{align}
		F(x_0)=&\left[F_1(x_0) + \beta F_2(x_0) \right] \Theta(x-x_0) \nonumber \\
		&+  \left[F_3(x_0) + \gamma F_4(x_0) \right] \Theta(x_0-x)
	\end{align} 
	where $\Theta$ stands for the Heaviside step function. Enforcing the matching conditions $F(x^+)-F(x^-)=0$ and $F'(x^+)-F'(x^-)=\varepsilon F(x)$, one obtains
	\begin{align}
		\beta&=\frac{[F_3-F_1]G_4+F_4[G_1-G_3+\varepsilon F_1]}{F_2G_4-F_4[G_2+\varepsilon F_2]}, \\
		\gamma&=\frac{[F_3-F_1]G_2+F_2[G_1-G_3+\varepsilon F_3]}{F_2G_4-F_4[G_2+\varepsilon F_2]},
	\end{align}
	where we omit from now on the argument $(x)$ in $F$'s and $G$'s.
	Now, we can evaluate $\tilde{F}(x_T,x)$
	\begin{align}
		\tilde{F}(x_T,x)=&F_3(x_T) \nonumber \\
		&+\frac{[F_3-F_1]G_2+F_2[G_1-G_3+\varepsilon F_3]}{F_2G_4-F_4[G_2+\varepsilon F_2]} F_4(x_T).
	\end{align}
	Carrying out the derivative with respect to $\varepsilon$ and evaluating at $\varepsilon=0$, we get
	\begin{align}
		M(x_T,x)=&-\frac{F_2F_4[G_1-G_3]+F_2F_3G_4-F_1F_4G_2 }{\left[F_4G_2-F_2G_4\right]^2} F_2 \nonumber \\ & \quad \times F_4(x_T).
	\end{align}
	
	Comments on the need of building objects for our numerical computation:
	\begin{itemize}
		\item For Case II, we need $U_{12}(x_T,x)$ for the region $0<x_T<x<1$, which actually implies the need of $U_{22}(x_T,x)$ too. These components are decoupled from $U_{11}$ and $U_{21}$, which are not needed. Additionally, $F_1$ and $F_2$ for $0<x<1$ are required, which implies the solution for $F_1$ and $F_2$ for the whole interval $x_0 \in (-1,1)$.
		\item For Case III, we need $F_1$, $F_2$, $F_3$, $F_4$, $G_1$, $G_2$, $G_3$, and $G_4$, for $x_0 \in (-1,0)$. Furthermore, we need $F_4$ for $x_0 \in (0,1)$.
	\end{itemize}
	
	\section{Robustness of numerical results for other target distributions}
	\label{ap:rob}
	
	The convergence and accuracy of the numerical optimization procedure is a subtle issue.  For other target distributions, different from the beta family considered in Eq.~\eqref{eq:PTbeta}, we have obtained qualitatively similar results, as shown below. 
	
	For example, we have also considered the case of a simple polynomial distribution:
	\begin{equation}
		P_{\text{pol}}^{(n)}(x_T;b)=a_n(b) + b|x|^n.
		\label{SMeq:poly1}
	\end{equation}
	We have explored this family, mainly for $n=1,2$. The coefficient $a_n$ is chosen to guarantee normalization of the distribution. The sign of $b$ makes the probability accumulate either around the center, for $b<0$, or close to the walls, for $b>0$. 
	
\begin{figure}
		\centering
		\includegraphics[width = 0.49\textwidth]{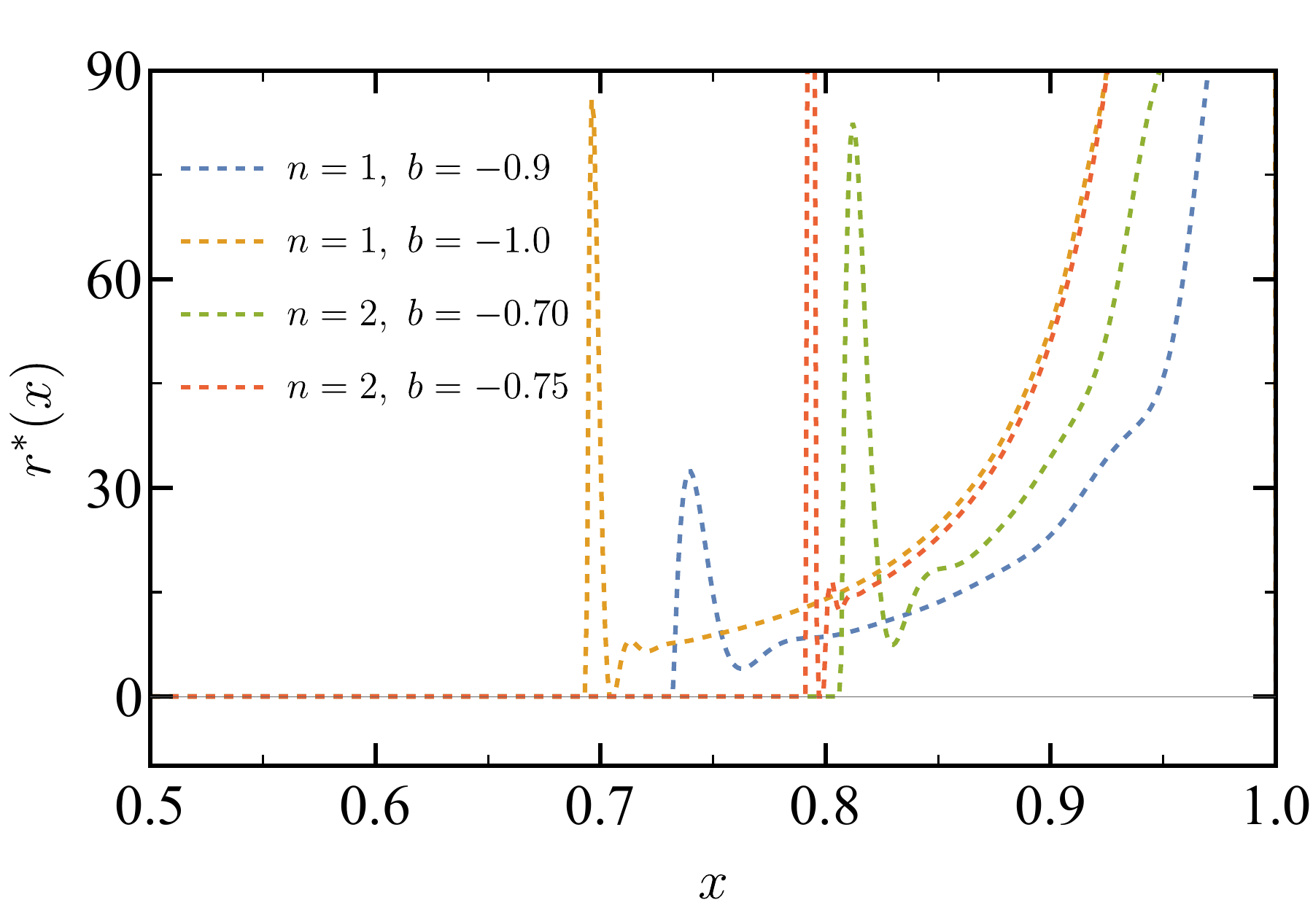}
		\caption{Optimal strategy for the polynomial family in Eq.~\eqref{SMeq:poly1} for $N=2001$.}
		\label{figSM:poly1op}
	\end{figure}

	Using a completely analogous procedure to that discussed in the main text, one  shows that there is a transition at a critical value of $b$, $b_c$, which separates regions of $b$-values for which the optimal resetting strategy in the bulk vanishes or not. For optimal nonzero resetting strategy, the numerically obtained solution shares the features displayed in the main text for the beta family---a nonresetting spatial window close to the center, followed by a peak and a strongly heterogeneous resetting profile when getting close to the walls. This hints at the existence of common features in the optimal resetting strategies, when the distribution is made by just one ``hill'' centered at the origin. The numerically obtained optimal resetting strategies are shown in Fig.~\ref{figSM:poly1op}, for different values of the parameters in the polynomial family~\eqref{SMeq:poly1}~\footnote{For some of them, the advantage of the change in the strategy is so low that the numerical optimization gets frozen. This is especially sensible close to the walls, where the impact of the change in the resetting strategy is very low, due to the smallness of $P$ in that region.}.
	
	\section{An heuristic interpretation of the Dirac delta resetting}
	\label{ap:dirac}
	
	The introduction of a Dirac delta peak in a resetting profile,
	\begin{equation}
		r(x) = R \, \delta (x - x_R) \ ,
	\end{equation}
	
	is well-defined at the Fokker-Planck equation level in Eq.~\eqref{SMeq:forward1}. Its practical implementation and physical interpretation can be understood with an heuristic argument. 
	
	In a computer simulation, a displacement $\Delta x_t \sim \sqrt{D\Delta t}$ is proposed for the searcher at time $t$, and it is rejected in case of resetting. The latter can occur with amplitude $R$ within a narrow interval around $x_R$:
	\begin{equation}
		r(x) = 
		\begin{cases}
			R/\varepsilon & \vert x-x_R \vert < \varepsilon/2 \\
			0 &\text{otherwise}
		\end{cases}
		\ ,
	\end{equation}
	
	If $\varepsilon \ll \sqrt{D \Delta t}$, the increment $\Delta x_t$ will likely displace the searcher across the resetting window $[x_R-\varepsilon/2,x_R+\varepsilon/2]$. A linear interpolation of the trajectory allows one to compute the time spent in the resetting interval as $\Delta t_R = \varepsilon / \vert v_t \vert = \varepsilon / \vert \Delta x_t / \Delta t \vert$.
	
	The searcher reaches the end of the resetting interval if no resetting occurs in the time $\Delta t_R$; the probability of this event can be estimated as
	
	\begin{equation}
		\label{SM:dirac-prob}
		\text{Prob (no resetting)} = e^{-(R/\varepsilon) \Delta t_R} = e^{- R / \vert v_t \vert} \ .    
	\end{equation}
	
	In the $\varepsilon\to0$ limit, this result can be interpreted as the probability of 
	not
	being reset every time the searcher crosses a Dirac delta resetting point $x_R$. Numerically speaking, every time the searcher diffuses across $x=x_R$ with a space increment of $\Delta x_t$, the move is accepted with a probability given by Eq.~\eqref{SM:dirac-prob}. Therefore, the resetting probability is determined by the searcher's speed at $x=x_R$: faster searchers have a lower probability of resetting because of their lower residence time around $x_R$.
	
	As one can immediately argue, the resetting probability during a single crossing vanishes in the continuum limit where $\vert v_t\vert \sim \sqrt{D/\Delta t} \to \infty$:
	\begin{equation}
		\text{Prob(resetting}) = 1 - e^{- R / \vert v_t \vert} \sim R \sqrt{\Delta t/D} \ .
	\end{equation}
	
	However, in the same limit $\Delta t\to 0$ the searcher is wiggling infinitely fast across the resetting point $x_R$. The number of crossings of the resetting point scales as $N_c \sim 1/\sqrt{\Delta t}$~\cite{redner_book}. The resetting probability per single crossing, multiplied by the characteristic number of crossings, thus converges to a finite value, and resetting survives in the continuum limit. 
	
	\bibliography{bibl}
\end{document}